\documentstyle[12pt]{article}

\begin{document}
\sf

\input psfig


\hbox{}

\centerline {\large \bf North Atlantic thermohaline circulation }
\centerline {\large \bf predictability in a coupled ocean-atmosphere model}

\vskip 1.5truecm

\centerline{Stephen M. Griffies{\dag} and Kirk Bryan{\ddag}}

\vskip .50truecm

\centerline{Princeton University}
\centerline{Atmospheric and Oceanic Sciences Program}
\centerline{Sayre Hall, Forrestal Campus}
\centerline{Princeton University, Princeton NJ 08544-0710}

\vskip .50truecm
\centerline{{\dag}email: smg@gfdl.gov}
\centerline{{\ddag}email: kbryan@splash.princeton.edu}

\vskip 2.5truecm

\centerline{Submitted to {\em Journal of Climate}}
\centerline{Los Alamos e-print ao-sci/9502001}

\vskip 1.0truecm

\vfill\eject


\centerline{\large \bf Abstract}

Predictability of the North Atlantic thermohaline circulation (THC)
variability as simulated in the Geophysical Fluid Dynamics
Laboratory's coupled ocean-atmosphere general circulation model is
established for a set of ensemble experiments.  There is a large
separation of time scales between the slower oceanic processes, whose
predictability is of interest here, and the much more rapid
atmospheric processes to which the ocean is coupled.  The ensembles
consist of identical oceanic initial conditions underneath a model
atmosphere chosen randomly from the model climatology.  This
experimental design is based on the separation in time scales present
in the model which motivates the assumption that the predictability
deduced from these ensembles provides an upper limit to the model's
THC predictability.  The climatology, against which the ensemble
statistics are compared, is taken from a multi-century model
integration whose THC variability has power concentrated at the 40-60
year time scale.  A linear stochastic perspective, motivated from
Brownian motion and THC box model case studies, is shown to be
generally consistent with the ensemble statistics.  The linear theory
suggests a natural measure of ensemble predictability as the time at
which the ensemble variance becomes a subjectively defined fraction
($50 \%$ used here) of the climatological variance.  It is furthermore
of interest to distinguish predictability of the rapidly
de-correlating portion of the model's THC from the longer time
correlated portion.  The rapidly de-correlating portion shows
predictability for $\approx 1.5$ years.  The slower portion shows
predictability for $\approx 5-7$ years.  The linear stochastic
framework provides for the straightforward construction of an optimal
forecast of the model's THC variability.  It also allows for an
understanding of why the optimal forecast is useless beyond a
particular predictability limit.

\vfill\eject

\section{Introduction}
\label{section:intro}

\subsection{North Atlantic variability}
\label{subsection:atlantic_variability}

Recent analysis of historical data by Levitus (1989a,1989b,1990),
Deser and Blackmon (1993), and Kushnir (1994) indicate the existence
of a significant amount of interannual to interdecadal variability in
the North Atlantic climate system.  The correlation between North
Atlantic climate variability and that of adjacent land regions is of
interest in part because of the desire to forecast that variability
most affecting human populations.  At present, there is insufficient
observational and proxy evidence to clearly indicate the significance
of the correlations seen in the data record.  For example, the
analysis by Broecker et al.\ (1985) of the large climatic fluctuation
from the warm Aller\"{o}d to glacial Younger Dryas periods
(approximately 11,000 years ago) suggests that North Atlantic
variability at that time was correlated to only certain regions of
western Eurasia and coastal regions of northeast North America into
Greenland.  The recent analysis by Schlesinger and Ramankutty (1994)
of the last 100 years of global surface air temperatures, however,
suggests a stronger and more coherent relationship between the North
Atlantic variability during this recent period and that of adjacent
land regions.  Yet it cannot be determined whether this behavior is
typical until longer time series of proxy data from different parts of
the globe become available.

A primary objective of several climate research programs, such as the
Atlantic Climate Change Program (ACCP), the Atlantic portion of the
World Ocean Circulation Experiment (WOCE), and the second phase of the
Climate Variability and Prediction Research Programme (CLIVAR), is to
design practical programs for monitoring climate variability of
decadal and multidecadal time scales such as that having been observed
in the North Atlantic. The monitoring would be focused on natural
climate variations and the detection of possible long term
anthropogenic climate change.  Presently, it is not clear what the
requirements for such a monitoring system are. One motivation for the
present study is to initiate studies useful for assessing the utility
of such an observing system for the prediction of natural climate
variability in the North Atlantic.  Furthermore, the extent to which
North Atlantic variability is correlated to that of surrounding land
regions may indicate the extent to which the monitoring of such
variability aids in forecasting climate variability of relevance to
human populations.  Presently, the TOGA TAO (Tropical Ocean and Global
Atmosphere / Tropical Atmosphere Ocean) array, in conjunction with
other measurements made from satellites and ships of opportunity,
provides input which appears to allow a useful basis for coupled
ocean-atmosphere model projections of El Ni\~{n}o -- Southern
Oscillation (ENSO) (e.g., Neelin et al. 1994). One question is whether
the very different air-sea interaction processes involved in North
Atlantic climate variability can likewise be projected forward in time
using coupled ocean-atmosphere models, but allowing for the added
involvement of the deep ocean in which the thermohaline circulation
(THC) plays an important role.

The THC is a major component of the ocean circulation and is
particularly important for poleward heat transport by ocean currents
to high latitudes in the North Atlantic.  This role of the ocean's THC
in affecting long term North Atlantic variability was suggested by
Bjerknes (1964) (see Bryan and Stouffer 1991 for a survey).
Therefore, assuming a close link between low frequency North Atlantic
climate variability and THC variability, the physical basis of North
Atlantic climate predictability is related to predictability of the
THC.  It is with this connection in mind that the current coupled
ocean-atmosphere model predictability study is undertaken with focus
on the model's THC variability.

\subsection{Modeling the thermohaline circulation}
\label{subsection:thermohaline-circulation}

Since the THC is thought to be especially important for North Atlantic
climate, numerous modeling efforts have focused on understanding the
stability and variability of the THC.  In regards to stability,
Stommel's two box model (1961) highlighted the importance of the
different oceanic boundary forcing contributed by heat and hydrology.
In this model, more than one stable mode of circulation is available:
one with the thermally dominant circulation corresponding to that
acting in today's climate and another with a reversed flow dominated
by haline effects.  F. Bryan (1986) initiated a continuing effort to
establish the stability of a hierarchy of ocean-only models forced
under mixed boundary conditions (linear damping of sea surface
temperature (SST) to a specified value and constant hydrological
fluxes affecting sea surface salinity (SSS)).  In a general
circulation model (GCM) simulation, he was able to create a shut down
of the THC after placing an anomaly of fresh water in the northern
portion of the model.  Further studies of ocean-only models using
mixed boundary conditions exhibit a wide range of stability and
variability properties depending on the strength of the forcing used
(see Weaver and Hughes 1992 for a survey).

Ocean-only climate models can be considered crude approximations to
the more physically complete coupled ocean-atmosphere models since the
surface boundary conditions in the ocean-only models are intended to
emulate coupling to an atmosphere.  Furthermore, as the THC is driven
by surface buoyancy fluxes embodied in the model's boundary
conditions, the behaviour of the THC in ocean-only models can vary
widely depending on the strength of the forcing and the feedbacks
incorporated.  Recently, some groups have begun to reexamine the
suitability of some mixed boundary condition models as an
approximation to the ocean-atmosphere coupling.  For example, Zhang et
al.\ (1993) and Mikolajewicz and Maier-Reimer (1994) have suggested
the relevance of particular modifications to their model's heat
transport feedbacks, as parametrized by surface temperature restoring
times, in such a manner which yields a more stable model behaviour.
Rahmstorf (1994) presents the results of a model forced with fixed
hydrological fluxes yet with a modified temperature boundary condition
acting in a scale dependent manner.  This model appears to be more
stable to an overall collapse of the THC under hydrological
perturbations although it is quite unstable to changes in the
particular convection pattern relevant for the model's THC.  In
addition to questioning the form of the temperature feedbacks acting
in the models, Tziperman et al. (1994) have questioned the use of
strong hydrological forcing which, under the usual mixed boundary
conditions, can result in highly variable and unstable model THC
simulations.  It is suggested that a more reasonable magnitude of
hydrological forcing results in a more stable mixed boundary condition
model.

In addition to formulating the deterministic boundary conditions
parametrizing the ocean-atmosphere interactions, understanding the
effects of atmospheric fluctuations acting on the ocean's surface can
be important for modeling the ocean's variability.  To date, these
fluctuations in ocean-only models have taken the form of a stochastic
forcing such as that used by the Max Planck Institute in Hamburg
(e.g., Hasselmann 1982, Mikolajewicz and Maier-Reimer 1990, Weisse et
al.\ 1994).  As discussed in the next subsection, we believe that a
stochastic perspective is appropriate for interpreting the coupled
model THC ensemble experiments reported in this paper.  It should be
noted that the precise form of a prescribed stochastic forcing (i.e.,
the spatial and temporal correlations, power of the stochastic
forcing, and feedbacks) acting on an ocean-only model inevitably plays
a large part in determining the model's response. Finding appropriate
forms of this non-deterministic forcing, which conceivably requires
the use of fully coupled ocean-atmosphere models as well as
observational data, is arguably at least as difficult a problem as
determining appropriate deterministic boundary conditions.

An alternative to highly parameterized representations of an atmosphere
coupled to a more detailed ocean circulation model is to couple a more
explicit numerical model of the atmosphere to an ocean model.  Such
models do not rely on the approximations to air-sea interactions
necessary for setting upper ocean boundary conditions in ocean-only
models.  Nor do they assume {\em a priori} a particular statistical
character of the atmospheric component.  Yet these models have their
problems as well.  Firstly, they are quite expensive to run.  For
example, the coupled model used in the present study requires 1.5--2.0
Cray-YMP hours per model year.  This expense greatly constrains the
statistics available from the coupled model ensemble experiments
reported here.  It furthermore motivates the study of simpler models
in combination with the more realistic models.  Secondly, fully
coupled models require adjustments of the ocean-atmospheric fluxes
(Manabe and Stouffer 1988, Sausen et al. 1988) in order to prevent a
drift of the model's mean state away from geophysically relevant
regimes.  The working hypothesis employed in the flux adjusted models
is that such adjustments, as they have no variability on the time
scales of interest, will not unrealistically affect the model's long
term variability.  There is no rigorous proof of this hypothesis other
than direct validation of the model output with the observational data
record.  The physical relevance of studies using models with
unrealistic mean states, which result from the non-flux-adjusted
coupled models used to date, is unclear.  From our perspective,
sufficiently validated flux adjusted coupled GCMs are the best
available tool for addressing North Atlantic climate variability
issues, including its predictability.

An example of a flux adjusted coupled model which exhibits climate
variability consistent with certain North Atlantic observations is
that of Delworth et al.\ (1993; henceforth DMS93) who used the
Geophysical Fluid Dynamics Laboratory's (GFDL) coupled global
ocean-atmosphere GCM.  This model presents a realistic simulation of
the air-sea interaction over the Atlantic basin on the yearly to
interdecadal time scales and produces a low frequency variability
qualitatively similar to that seen in the analysis of Levitus (1989a)
and Kushnir (1994).  Since the low frequency variability in the
model's North Atlantic is associated with THC fluctuations, the model
successfully embodies Bjerknes' hypothesis for North Atlantic
variability.  As can be seen in the model analysis of Manabe and
Stouffer (1988), which discuss the model's behaviour when the North
Atlantic THC is both active and shut down, the correlation between the
model's North Atlantic SST variability and adjacent surface
temperature over land masses suggests a distinct downstream influence
in the direction of atmospheric flow as found by Broecker et al.\
(1985) from paleoclimate data.  On the other hand, Schlesinger and
Ramankutty (1994) show that the single 50--60 year swing in climate
which shows up in historical records indicates an association of SST
in the North Atlantic with both upstream and downstream surface
temperature over the adjacent land areas.

Given the compelling model and observational evidence for low
frequency variability of the North Atlantic, as well as the important
role played by the ocean's THC, understanding the predictability of
model simulated THC is a necessary step towards assessing the
possibilities of producing long term North Atlantic climate forecasts.
It is for this reason that a study of the North Atlantic THC
predictability using the GFDL coupled ocean-atmosphere model has been
undertaken.

\subsection{A stochastic perspective}
\label{subsection:stochastic-perspective}

Climate systems, through the presence of various coupled sub-systems,
typically contain a broad range of space and time scales which can be
defined by representative auto-correlation scales.  In the context of
the coupled ocean-atmosphere system relevant for this study, time
scales range roughly from the short time scale synoptic atmospheric
processes (on the order of days) to the long time scale oceanic
processes (on the order of years and longer). It is the response of
the ocean on longer time and larger space scales through processes
associated with the THC that are of interest in the current study.

Atmospheric model and observational studies indicate a loss of
deterministic predictability of typical mid-latitude synoptic scale
motions at the 10 day time scale (see Lorenz 1969 for an early review
of atmospheric predictability).  At times longer than the synoptic
predictability time the atmosphere is effectively non-deterministic.
The recent work by Kleeman and Power (1994) and Power et al.\ (1995)
have pointed towards the importance of this random variability in
affecting the predictability of the coupled ocean-atmosphere system.
Nevertheless, the hope is that by exploiting the memory of the slower
time scale processes, namely the ocean circulation, useful long term
predictions of the coupled system are possible.  For example, it is
this oceanic memory which allows for better than climatology ENSO
forecasts up to a year in advance (e.g., Neelin et al. 1994).  Given
that the variability in the North Atlantic circulation affects the
long term atmospheric statistics in this region, it is necessary,
although not sufficient, to characterize the predictability limits of
the oceanic sub-system in order to assess the feasibility of
forecasting low frequency climate statistics.

Hasselmann (1976), through exploiting the time scale separation
between various portions of the climate system, demonstrated the
relevance of linear stochastic climate models for understanding the
red noise response of certain climate processes, such as SST, to white
noise forcing which can approximate that provided by rapidly
de-correlating portions of the system, such as synoptic atmospheric
variability.  Subsequent modeling efforts have supported and extended
Hasselmann's perspective.  For example, Mikolajewicz and Maier-Reimer
(1990) and Weisse et al. (1994) forced the Hamburg large scale
geostrophic ocean model (Hasselmann 1982) with spatially correlated
white noise fresh water fluxes.  These studies identified linear
oceanic modes of variability in the model's Atlantic basin.  A similar
study was carried out by Mysak et al. (1993) using a two dimensional
THC model with random hydrological fluxes.  This study also resulted
in the excitation of various oceanic modes.  From a more idealized
perspective, Bryan and Hansen (1994) considered the variability of the
Stommel (1961) two-box model when forced with white noise buoyancy
fluxes.  Their motivation was to extend Hasselmann's SST model to
include salinity effects in the process of producing a toy model
description of the variability seen in DMS93.  As an extension of
Bryan and Hansen, the study by Griffies and Tziperman (1995; hereafter
GT95) show how random heating of a mixed boundary condition four-box
model emulates the variability and temperature--salinity--THC phase
relations seen in the coupled model simulation of DMS93.  The
excitation of a linear damped oscillatory THC mode dominates the
variability in their four-box model.

\subsection{The present ensemble predictability study}
\label{subsection:problem}

Predictability studies are concerned with the feasibility of making
predictions for dynamical systems on the basis of incomplete or
``imperfect'' initial conditions.  Expressed in another way,
predictability is concerned with the length of time for which a
prediction can be carried out before errors in the initial conditions
rise to the values of normal variability.  From a modeling
perspective, predictability studies are a first step towards assessing
the possibilities of using the model to forecast the variability of a
particular phenomenon.  In the early stages of such studies, such as
the present, validation of the forecast against reality is not
considered.  Rather, a characterization of the model's precision is
useful in order to determine the significance of a potential forecast.
A tool for addressing the precision question is the ensemble
experiment in which the model is integrated a number of times starting
from slightly different initial conditions.  This method is widely
used in atmospheric model predictability studies (e.g., Lorenz 1969).
The model is exhibiting predictable behaviour at a particular time if
the ensemble statistics are distinguishable, to within a subjective
statistical criteria, from the statistics of a climatology.  For
linear Gaussian processes, which form the basis of the analysis in
this paper, a complete comparison of the ensemble and climatology
statistics is provided by comparison of their mean and variance.  (The
recent work of Anderson and Stern 1995 and references therein provide
additional discussion of these points.)

This paper considers predictability of the yearly averaged and 10 year
low pass yearly averaged amplitude of the THC simulated in the GFDL
coupled ocean-atmosphere GCM.  Comparison of the statistics from a
series of ensemble experiments and a multi-century integration provide
a characterization of the ensemble's predictability.  In the coupled
model context, ensemble predictability studies have also been carried
out mostly with ENSO being the phenomenon of interest (e.g., Goswami
and Shukla 1991).  The authors are unaware of similar studies being
conducted with focus on the North Atlantic THC.

A generic perturbation of the atmosphere typically translates within a
few days into a completely different atmosphere within the
climatology.  This behaviour motivates our taking the working
hypothesis that the predictability of long term oceanic variability is
not significantly affected by the precise initial condition of the
atmosphere.  Therefore, we consider here coupled model ensembles in
which the initial conditions for the atmosphere are taken randomly
from model climatology whereas the ocean's initial state is unchanged.
Based on the separation in time scales between the atmosphere and
oceanic variability considered here, THC predictability deduced from
these ensembles should serve as an upper limit to the model's THC
predictability.  Although the initial condition of the real ocean can
never be perfectly known, which implies that ensembles whose elements
have different oceanic initial conditions will be necessary to more
completely characterize the model's THC predictability, the
predictability time scales established here will serve as a useful
benchmark to which more realistic studies can be compared.

The remainder of this paper consists of the following sections.
Section \ref{section:coupled-model} summarizes the salient features of
the coupled model and its variability.  Thereafter, a description of
the ensemble experiments is presented.  Before analyzing the ensemble
statistics, we analyze ensemble experiments from a hierarchy of
simpler models building on a linear stochastic perspective.  For this
purpose, Section \ref{section:linear-systems} gives a discussion of
ensemble experiments generated by two low order linear stochastic
models; namely, the Brownian (or red noise) and harmonic Brownian
processes.  Section \ref{section:box-models} shows how the linear
stochastic analysis is useful for understanding the statistics from
the THC four-box model of GT95.  The presentation given in these two
sections attempts to establish a framework providing a concise
description of the coupled model ensemble statistics.  Furthermore, as
we are limited to rather small ensemble sizes with the coupled model
due to their expense, the ensembles from the low order models will be
presented with comparable sizes in order gain a sense of the extent to
which sampling errors are an issue.  Conclusions from these case
studies have been verified by ensembles many times larger.  The
lessons learned from the low order models are applied in Section
\ref{section:coupled-linear} to the coupled model ensembles.  Section
\ref{section:discussion-conclusions} presents a discussion of the
results and conclusions.  Appendix \ref{appendix:stationary_stats}
supplements Section \ref{section:linear-systems} by providing formulae
for the stationary statistics of the two Brownian processes.


\section{The coupled model ensemble experiments}
\label{section:coupled-model}

\subsection{The model}
\label{subsection:coupled-model}

The coupled ocean-atmosphere model used here is that of the GFDL and
has been employed extensively for global warming research.  We use the
same configuration as DMS93, in which the North Atlantic variability
was analyzed from a multi-century model integration.  Reference should
also be made to Stouffer et al. (1989) for information regarding the
model structure and performance as well as Manabe and Stouffer (1988,
1995), Manabe et al.\ (1991,1992), and Stouffer et al.\ (1994) and for
added details and analysis.

Briefly, the model's atmospheric component is an R15 spectral model in
the horizontal with 9 vertical levels.  The ocean component is a
global 12 level Bryan-Cox-Semter model with $3.75^{\circ}$ longitude
by $4.5^{\circ}$ latitude resolution.  Ocean mixing occurs with an
isopycnal scheme described by Tziperman and Bryan (1993) in addition
to horizontal and vertical background eddy diffusion on top of a
vertical convective adjustment.  The model contains the seasonal cycle
and employs monthly and geographically dependent adjustments of the
heat and fresh water fluxes (Manabe and Stouffer, 1988).

\subsection{The model THC variability}
\label{subsection:model-variability}

Figure \ref{fig:climate_400years} shows 400 years of linearly
detrended (i.e., a least squares fit of a line to the time series has
been subtracted) anomalous yearly averaged THC index from the central
part of a 1000 year coupled model experiment (provided by Ron
Stouffer; see DMS93 and Manabe and Stouffer 1995 for analyses of this
model integration).  It is this portion of the millennium which serves
as a climatology for the subsequent ensemble experiments.  The THC
index is defined as the maximum value for the North Atlantic
meridional circulation streamfunction within the latitudes
$40.5\mbox{}^{\circ} N$ to $72\mbox{}^{\circ} N$.  It is a useful
index for measuring the strength of the THC in the North Atlantic
region of the model.  Furthermore, it is a spatially integrated
quantity that is useful for capturing the low frequency variability of
the model ocean, which is also associated with low frequency model
atmospheric variability (DMS93).  To further highlight the low
frequency variability, a 10 year low pass filtered version of the
time series is indicated.

For the purposes of conducting a predictability study, a stationary
climatology whose variability corresponds to that seen in the ensemble
experiments is necessary to quantify the ensemble's predictability.
We have chosen the 400 years shown in Figure
\ref{fig:climate_400years} based on this requirement and the added
desire to have a simple linear detrend produce stationarity.

Figure \ref{fig:climate_400years_acf} shows the auto-correlation
function ({\em acf}) of the time series in Figure
\ref{fig:climate_400years}.  Fit (using a least squares criterion) to
the {\em acf} are the {\em acf}s from the Brownian or red noise
process (fit to lags $\le 10$ years) and that from a noise driven
damped harmonic oscillator, also known as the harmonic Brownian
process (fit to lags $\ge 10$ years).  These linear stochastic
processes are the continuous versions of first and second order
auto-regressive processes, respectively (Jenkins and Watts, 1968) and
will be described further in Section \ref{section:linear-systems} and
Appendix \ref{appendix:stationary_stats}.  The Brownian process
captures the small time lag de-correlation of the THC index whereas
the harmonic Brownian process captures the negative trough suggesting
a damped oscillatory component in the THC. Similar fits result from
using slightly different lag regions.

The above fit was also done to the {\em acf} from the low pass
filtered signal (not shown) and the time scales are given in the
caption to Figure \ref{fig:climate_400years_acf}.  As described by
Munk (1960), filtering increases the auto-correlation of a damped
process, such as the Brownian process, since it filters out part of
the rapidly de-correlating portion of the process.  Hence, the
correlation e-folding times are larger for the filtered process.

It should be noted that the statistical significance of the negative
trough in the {\em acf} of Figure \ref{fig:climate_400years_acf} is
marginal at the $95 \%$ confidence level.  Analysis of different
portions of the climatology (DMS93 and GT95), as well as certain
fields [e.g., dynamic topography (not shown)], nevertheless motivates
us to consider the climatology to have a highly damped low frequency
oscillatory component coupled to purely damped components, rather than
being the result of completely damped processes.  This interpretation,
which is perhaps a minimal description of the coupled model's THC
variability, is also motivated by the box model study of GT95 and
discussed in Section \ref{section:box-models}.  The spectral analysis
of DMS93 indicate that most of the THC power is broadly spread about
the 40-60 year time range.  This breadth corresponds to the broad
trough in the {\em acf}s relative to that of the single damped
harmonic oscillator fit in Figure \ref{fig:climate_400years_acf}.

\subsection{The ensemble experiments}
\label{subsection:coupled-ensembles}

Two sets of ensemble experiments are reported in this paper and are
shown in Figures \ref{fig:ensI} and \ref{fig:ensII}.  The elements of
the first ensemble was started from year 130 of the climate run and
the second ensemble was started at year 500.  Both ensembles extend
for 30 years and both started with an initial anomalous circulation
greater than one standard deviation from the mean.

The stochastic nature of these time series suggests the hypothesis
that the model's THC variability is associated with the effectively
random forcing from the model's atmosphere.  This idea was suggested
in the discussion of DMS93, studied in a box model context by Bryan
and Hansen (1993) and GT95, and will be pursued further in the
following sections.

\section{Predictability in linear noise driven systems}
\label{section:linear-systems}

In this section, we discuss two linear, single variable, stochastic
models in which a complete statistical description is available from
concise and explicit mathematical formulae.  Analysis of the
predictability of these models' variability will be useful for
establishing the basics of ensemble predictability experiments.
Furthermore, this analysis will prove to be a relevant general
framework for understanding the statistics from the coupled model
ensemble experiments.

\subsection{The Brownian process}
\label{subsection:brownian}

As our first example of linear noise driven processes, consider the
{\em red noise} or {\em Brownian process} $v(t)$ described
by the equation
 \begin{equation} \dot{v}(t) = - \alpha v(t) + \xi(t),
 \label{eq:red}
 \end{equation}
 where the overdot indicates a time derivative.  The parameter $\alpha
\ge 0$ represents the effects of frictional forces which are
dissipative and act to relax $v(t)$ back to its equilibrium value
$v_{eq}(t) = 0$; i.e., it provides a linear negative feedback.
$\xi(t)$ represents the rapidly de-correlating forces.  This is a
fluctuation term which will be modeled as a Gaussian white noise
process with zero mean and auto-covariance $\langle \xi(t) \xi(s)
\rangle = \sigma_{\xi}^{2} \delta(t-s)$, with $\delta(t)$ the Dirac
delta function and $\sigma_{\xi}^{2}$ determining the power of the
noise.  The expectation operator $\langle \bullet \rangle$ represents
an average over an infinite sized ensemble.  As the stochastic
processes considered in this paper are linear Gaussian, the first
moment stationary statistics from a finite sized ensemble will
approximate the statistics of the infinite sized ensemble with
correction terms proportional to $N^{-1/2}$, where $N$ is the ensemble
size.

 In the context of Brownian motion, $v(t)$ represents the velocity
$\dot{x}(t)$ of the Brownian particle (see e.g., Reif 1965).
Hasselmann (1976) introduced this process into the climate literature
as a means of describing the integrative response of a slow climate
sub-system, such as the ocean, to a shorter time scale forcing, such
as that provided by the synoptic scale atmosphere.  We will
generically refer to this equation as that describing a {\em Brownian
process}.  It is this process which will be argued to be useful for
understanding the behaviour of the relatively short time scale portion
of the coupled model THC variability.  Some details regarding the
stationary statistics of this process are given in Appendix
\ref{appendix:stationary_stats}.

\subsubsection{The optimal forecast}
\label{subsubsection:optimal_red}

Although the Brownian process is stochastic and hence
non-deterministic, it might still be of interest to produce an optimal
forecast of a future state.  For this purpose, given observations of a
particular realization $v(t)$ over some time ending at $t=0$, it is
useful to ask what is the most probable future state at times $t =
\tau > 0$?  As discussed recently by Penland (1989) in the
meteorological context, since the process is Gaussian, the most
probable future state of the system is also that state which is
optimal in a least squares sense.  For the Brownian process, this
state defines the {\em damped persistence} forecast (see Lorentz 1973
for the discrete case)
 \begin{equation}
 v_{dp}(\tau) = v(0) \exp(-\alpha \tau);
 \label{eq:red_forecast}
 \end{equation}
 i.e., this forecast is a damping of a persistence forecast
$v_{persist}(\tau) = v(0)$ back to a zero anomaly with a relaxation
time given by the system's auto-correlation time.  It is useful to
note that this forecast is the same as the mean of an infinite sized
ensemble each of whose elements starts at the initial state $v(0)$.

Since each ensemble element starts from the initial state $v(0)$, a
formal expression for a particular ensemble element can be written
 \begin{equation}
 v(\tau) = v(0)e^{-\alpha \tau} +
           \int^{\tau}_{0} e^{-\alpha(\tau-u)} \xi(u) du \equiv
           v_{dp}(\tau) +  \int^{\tau}_{0} e^{-\alpha(\tau-u)} \xi(u) du,
 \label{eq:solution}
 \end{equation}
 where the white noise process $\xi(u)$ differs for each element and
the damped persistence forecast (\ref{eq:red_forecast}) was
identified.  Hence, the mean square difference between the forecast
$v_{dp}(\tau)$ and an infinite number of realizations of the process,
each of which starts at $v(0)$, provides a natural definition of the
error in the forecast:
 \begin{equation}
 \sigma_{dp}^{2}(\tau) \equiv  \langle (v(\tau) - v_{dp}(\tau))^{2}\rangle
   =  \langle v^{2} \rangle (1 - e^{-2 \alpha \tau}),
 \label{eq:red_sigma}
 \end{equation}
 where $\langle v^{2}\rangle =\sigma_{\xi}^{2} / 2 \alpha$ is the
climatological or stationary variance of the process (see Appendix
\ref{appendix:stationary_stats}).  The error is also understood as the
variance in the infinite sized ensemble about its mean state as well
as the covariance between the forecast and an infinite number of
realizations (again, each of which starts at $v(0)$).  It is this
measure of the forecast error, and its approximation from finite
ensembles, which is appropriate for measuring the predictability
remaining in an ensemble forecast of linear Gaussian noise driven
systems.

The squared forecast error $\sigma_{dp}^{2}$ approaches the
climatological variance with an e-folding time $1/(2\alpha)$.  The
larger the feedback coefficient $\alpha$ of the Brownian process, the
faster the forecast error saturates to that defined by the
climatology.  Note that this e-folding time for the saturation of the
forecast error is one-half the auto-correlation e-folding time for the
process.  The same error calculation for the persistence forecast
$v_{persist} = v(0)$ yields the mean squared error
 \begin{equation}
 \sigma_{persist}^{2}(\tau) = \langle (v(\tau) - v(0))^{2} \rangle =
 2 \, \langle v^{2}\rangle (1 - e^{-\alpha \tau}),
 \label{eq:persist}
 \end{equation}
 which saturates to twice the climatological variance yet at a slower
rate than the optimal forecast.  The error in the persistence forecast
is always larger than that of damped persistence.  The slope of the
optimal forecast error is given by
 \begin{equation}
  {d \sigma^{2}_{dp} \over dt} = 2\alpha \langle v^{2} \rangle
  e^{-2\alpha \tau},
  \label{eq:red_optimal_slope}
  \end{equation}
 which indicates an exponentially decaying but positive slope starting
at $2\alpha \langle v^{2} \rangle$. Filtering a Brownian process will
reduce this initial slope, as expected from there being an increase in
persistence for filtered signals (Munk 1960).

Some numbers corresponding to the coupled model THC index
auto-correlation time scales (Figure \ref{fig:climate_400years_acf})
are illustrative.  With a subjectively defined level of forecast skill
$\gamma \equiv \sigma_{dp}^{2}/\sigma_{climate}^{2} = [1-\exp(-2\alpha
\tau_{\gamma})]$, the time $\tau_{\gamma}$ beyond which the optimal
forecast is of no use, which serves to define a {\em predictability
time}, is given by $2\alpha \tau_{\gamma} = - \ln(1-\gamma)$.  For
example, with $\gamma = .5, 1/\alpha = 5$ years, $\tau_{\gamma} =
1/(2\alpha) \ln 2 = 1.7$ years.  For the less stringent level $\gamma
= .75$, the useful lead time is $\tau_{\gamma} = 3.5$ years.  The
equivalent calculation for the persistence forecast, whose error is
given by equation (\ref{eq:persist}), yields $\alpha \tau_{\gamma} =
-\ln(1-\gamma/2)$, or a $.5$ error crossing time of 1.4 years and a
.75 error crossing time of 2.4 years.

\subsubsection{A numerical example}
\label{subsubsection:numerical-brownian}

A finite set of ensemble elements may exhibit either slower or faster
saturation of ensemble variance than that described above (to within
the constraints set by the sampling errors proportional to $N^{-1/2}$,
where $N$ is the ensemble size).  Furthermore, as seen in the
experiments of this and the following sections, the finite ensemble
results most closely approximate the infinite ensemble results at
small lead times.  The agreement between the finite ensemble results
and the infinite results will also depend on the level of the noise in
the system, with a higher noise level resulting in more variance about
the infinite ensemble results.  These points are illustrated in
Figures \ref{fig:Brown_corr} and \ref{fig:Brown_ensemble}.  Figure
\ref{fig:Brown_corr}A shows a 200 year realization of the Brownian
process.  The parameters used for realizing this signal are indicated
in the caption.  The {\em acf} for 400 years of the process is shown
in Figure \ref{fig:Brown_corr}B.  Figure \ref{fig:Brown_ensemble}A
shows a nine element ensemble.  The noise forcing is identical for all
the elements up to year 65 and differs thereafter.  The ensemble mean
is shown in Figure \ref{fig:Brown_ensemble}B.  The smoother dashed
line is the mean of the infinite sized ensemble; i.e., the damped
persistence forecast given by equation (\ref{eq:red_forecast}).
Figure \ref{fig:Brown_ensemble}C shows the ensemble variance.  The
smoother dashed line is the expected variance (\ref{eq:red_sigma})
from the infinite sized ensemble; i.e., the optimal forecast error.
Note that we follow the discussion of Penland (1989) for numerically
realizing the stochastic processes considered in this paper (see
Kloeden and Platen 1992 for a mathematical treatment).

The above results form the basis of optimally forecasting a red noise
or Brownian process.  Once an estimation of the system's initial state
$v(0)$ (from observations) and auto-correlation time $\alpha^{-1}$
(from the climatology) has been made, the damped persistence forecast
(\ref{eq:red_forecast}), which exponentially damps an anomaly back to
zero, is available. This forecast is the best forecast possible in the
least squares sense for the red noise process and it is also the most
probable future state of the process.  The only means for making a
useful prediction is afforded by the system's memory.  The
effectiveness of the damped persistence forecast, as measured by its
predictability or ensemble variance e-folding time, is directly
related to the system's correlation time.

\subsection{The harmonic Brownian process}
\label{subsection:oscillator}

This section presents the analogous calculations of the previous
section for the damped harmonic oscillator driven by Gaussian white
noise.  This process will be referred to here as the {\em harmonic
Brownian process}.  It is this process which will be argued to be
useful for understanding the coupled model's THC low frequency
variability.  The equation describing this process is given by
 \begin{equation}
 \ddot{x}(t) + \omega_{0}^{2} x(t) = -2\beta \dot{x}(t) + \xi(t).
 \label{eq:oscillator}
 \end{equation}
 The parameter $\beta \ge 0$ represents the contribution of frictional
or viscous dissipation, the frequency $\omega_{0}$ is the natural
frequency of the noise free and dissipation free system ($ \xi(t) =
\beta = 0$), and the white noise forcing $\xi(t)$ is like that in the
Brownian process example.  Section
\ref{subsection:app_harmonic_brownian} in the Appendix discusses the
stationary statistics for this process.  The remainder of this section
describes the problem of forecasting and characterizing the
predictability of this system when it is underdamped; i.e.,
$\Omega^{2} \equiv \omega_{0}^{2} -\beta^{2} > 0$.

\subsubsection{The optimal forecast}
\label{subsubsection:dho_optimal}

 As in the last subsection, since the process is linear and Gaussian,
the optimal forecast (also the most probable future state) is the same
as the mean of an infinite ensemble; i.e., it is the noise free
solution of (\ref{eq:oscillator}) whose initial conditions are the
state and time tendency at the moment of the last measurement.  The
forecast satisfying these conditions is given by
 \begin{equation} x_{dhp}(\tau) = { e^{-\beta \tau} \over \Omega }
  \{ x(0) \Omega \cos(\Omega \tau) + [\dot{x}(0) +\beta
 x(0)]\sin(\Omega \tau)\}.
 \label{eq:ho_forecast2}
 \end{equation}

The optimal forecast (\ref{eq:ho_forecast2}) will be called {\em
damped harmonic persistence} in analogy with the damped persistence
forecast given by (\ref{eq:red_forecast}).  Note that for certain
initial conditions relevant for the growing phase of an oscillation,
the damped harmonic persistence forecast results in an initially
growing anomaly.  This growth is in contrast to the damped persistence
forecast (\ref{eq:red_forecast}) which always results in a damping of
anomalies and hence can never capture a growing anomaly.  Once the
initial state $x(0)$ and time tendency $\dot{x}(0)$ are measured, the
damped harmonic persistence forecast can be constructed after the
damping coefficient $\beta$ and natural frequency $\omega_{0}$ are
estimated from the {\em acf} of a climatology.

The mean squared error in the damped harmonic persistence forecast
$\sigma_{dhp}^{2} = \langle (x(\tau) - x_{dhp}(\tau))^{2}\rangle$
[which is also the ensemble variance where each ensemble element
starts with the same $x(0)$ and $\dot{x}(0)$] is given by
 \begin{equation}
 \sigma_{dhp}^{2}(\tau) =
 \langle x^{2} \rangle \{ 1 - { e^{-2 \beta \tau} \over \Omega^{2} }
 [ \omega_{0}^{2} -
 \beta^{2} \cos(2 \Omega \tau) + \beta \Omega \sin(2 \Omega \tau) ] \}.
\label{eq:ho_forecast_error}
\end{equation}
 The error $\sigma_{dhp}^{2}$ exponentially saturates to the
climatological variance $\langle x^{2} \rangle =
\sigma_{\xi}^{2}/(4\beta \omega_{0}^{2})$. The corresponding error for
the persistence forecast $x_{persist} = x(0)$ is
 \begin{equation}
 \sigma_{persist}^{2}(\tau) =
 2\, \langle x^{2} \rangle \{ 1 - e^{-\beta \tau}
  [ \cos(\Omega \tau) + {\beta \over \Omega} \sin(\Omega \tau)] \},
 \label{eq:persist-dho}
 \end{equation}
 which saturates to twice the climatological variance and is always
larger than the damped harmonic persistence forecast error.

Both forecast errors oscillate due to the oscillatory nature of the
process.  For the cases considered in this paper, this oscillation is
not so apparent because of the dominance of the exponential prefactor.
The slope of the optimal forecast error is given by
 \begin{equation}
 {d \sigma^{2}_{dhp} \over dt} = {4 \beta \omega_{0}^{2} \langle x^{2} \rangle
 \over \Omega^{2} } e^{-2\beta \tau} \sin^{2}(\Omega \tau),
 \label{eq:optimal_slope}
 \end{equation}
 which vanishes at zero lead time.  Recall the slope
(\ref{eq:red_optimal_slope}) for the error made in the optimal
forecast of the Brownian process (the damped persistence forecast) is
$2\alpha \langle v^{2} \rangle$ at zero lead time.

It is illustrative to compare the times at which the errors in the
damped harmonic persistence and pure persistence forecasts reach some
subjective fraction of the climatological variance.  For example,
consider a harmonic Brownian process of period 33 years and decay time
10 years, which correspond to values estimated from the coupled model
{\em acf} in Figure \ref{fig:climate_400years_acf}.  The error in the
damped harmonic persistence forecast (\ref{eq:ho_forecast_error}) of
this process takes 7.2 years to reach $50 \%$ of the climatological
variance and 9.7 years to reach $75 \%$.  The pure persistence
forecast error (\ref{eq:persist-dho}) reaches $50 \%$ after
4.4 years and $75 \%$ after 5.6 years.

\subsubsection{A numerical example}
\label{subsubsection:numerical_harmonic_brownian}

Figure \ref{fig:harm_Brown_corr}A shows a 200 year realization of the
harmonic Brownian process.  The model parameters are indicated in the
caption.  The {\em acf} from a 400 year realization is shown in Figure
\ref{fig:harm_Brown_corr}B.  Figure \ref{fig:harm_Brown_ensemble}A
shows a nine element ensemble.  The ensemble mean is shown in Figure
\ref{fig:harm_Brown_ensemble}B along with a damped harmonic
persistence fit given by equation (\ref{eq:ho_forecast2}).  Figure
\ref{fig:harm_Brown_ensemble}C shows the ensemble variance.  The
smoother dashed line is the variance (\ref{eq:ho_forecast_error}) from
the damped harmonic persistence forecast.

\section{Linear predictability in a THC box model}
\label{section:box-models}

\subsection{Introduction}
\label{subsection:box_intro}

In this section, the previous study of variability generated by
Brownian and harmonic Brownian processes will be used to help guide
our characterization of the variability seen in a white noise driven
THC box model.  Such a study provides the next level in a hierarchy of
models for studying the variability of the coupled model's THC. Box
models are low order systems which have proven to be of use for
understanding the stability and variability of the North Atlantic THC.
The model of interest here is an extension of the familiar advective
two-box model of Stommel (1961) to that of four boxes.  The focus in
this section, as in the preceding discussion, is on the variability of
the box model around some stable stationary state.  The evolution of
Stommel-type (Stommel 1961, Welander 1986) box models are determined
by conservation laws of salinity and heat applied to various well
mixed regions or boxes.  In their canonical form, the only
nonlinearity occurs in advection terms if the equation of state for
sea water is taken as a linear function of temperature and salinity:
$\rho_{i} = \rho_{0}[1 - \alpha_{T}(T_{i}-T_{0}) +
\beta_{S}(S_{i}-S_{0})]$ with $\rho_{0}, T_{0}$, and $S_{0}$ being
reference density, temperature, and salinity, respectively.  The
thermal and saline expansion coefficients $\alpha_{T}$ and
$\beta_{S}$, whose dimensions are $\mbox{}^{\circ}\mbox{K}^{-1}$ and
$\mbox{psu}^{-1},$ respectively, are taken as constants throughout the
model which guarantees that the advective circulation vanishes when
the north/south temperature and salinity gradients vanish.  Hence, the
circulation is driven solely by horizontal pressure gradients between
the north and south which, using the hydrostatic approximation,
correspond to horizontal density gradients.  More realistic nonlinear
equations of state are thought to give no qualitative change in the
model's behaviour under the moderate variability considered here.

GT95 argued for the ability of a particular two dimensional, mixed
boundary condition four-box model to emulate the coupled model's THC
variability and the zonally averaged circulation variability.  The
same box model, without stochastic forcing, was previously considered
by Huang et al.\ (1992) and Tziperman et al.\ (1994).  As shown in
GT95, it is the excitation of a damped oscillatory eigenmode by random
atmospheric forcing, modeled as white noise, which is important for
characterizing the low frequency box model variability.  Coupled to
this damped oscillatory mode are purely damped modes which are also of
relevance to establishing the model's predictability.

The experiments shown here are the result of numerically integrating
the nonlinear conservation equations given in GT95.  The numerical
scheme described in Penland (1989) is employed using 365 time steps
per model year.

\subsection{The four box model}
\label{subsection:four-box}

Figure \ref{fig:4box_geometry} shows the configuration of the four-box
model.  The model parameters, described further in GT95, are given in
the caption.  The temperature and salinity of each box are time
dependent hence providing eight degrees of freedom.

Linearization of the mixed boundary condition box model governing
equations about a stable thermally dominant steady state, realized
using the parameters given in Figure \ref{fig:4box_geometry} with
restoring boundary conditions on both temperature and salinity, yields
a damped oscillatory eigenmode of period $62$ years and decay time
$10$ years, respectively, as well as purely damped modes.  Since the
linear system is not self-adjoint, the eigenmodes are coupled to one
another.  The box model steady state, and its variability, were tuned
to emulate that seen in the coupled model.

Figure \ref{fig:bx4_signal} shows 400 years of yearly averaged four
box model circulation driven by a white noise surface heating.  It is
this time series which is used as a climatology necessary to quantify
predictability of the subsequent box model ensemble experiments. Also
shown is the signal after applying a 10 year low pass filter.  Low
passing highlights the response due to the low frequency damped
harmonic eigenmode as discussed in GT95.

Figure \ref{fig:bx4_signal_acf} shows the {\em acf} for the box
model's THC as well as fits by the {\em acf} for the Brownian and
harmonic Brownian processes.  Note the broad trough in the box model's
{\em acf} which is similar to that in {\em acf} of the coupled model
THC index (Figure \ref{fig:climate_400years_acf}).  As in the coupled
model, the statistical significance of the trough is marginal at the
$95 \%$ confidence level.  However, for the box model, a direct
analysis of the model's linearized dynamics supports our interpreting
the trough as a contribution to the model's variability from a damped
oscillatory THC eigenmode.  The {\em acf} from substantially longer
time series also supports this view.  It is straightforward to verify,
through box model experiments in which the oscillatory eigenmode is
either more or less damped, that the breadth and depth of the trough
is related to the rather large damping of the oscillatory eigenmode,
the presence of the purely damped eigenmodes to which the oscillatory
mode is coupled, and sampling errors.  The correspondence of the box
model's circulation statistics to that of the coupled model's THC
index, as well as the general correspondence between the zonally
averaged mechanisms acting in the two models as studied by GT95,
motivate a stochastic interpretation for the coupled model's THC
variability similar to that acting in the box model.

\subsection{Four-box model THC predictability}
\label{subsection:4box-first-predictability}

We now characterize the predictability of the THC variability
simulated by the stochastically forced box model.  In this analysis,
it will be of interest to quantify the predictability of the lower
frequency variability which is largely associated with the damped
oscillatory mode in addition to the predictability of the yearly
averaged variability.  For obtaining predictability of the low
frequency variability, it will be necessary to low pass filter the
ensemble elements in order to remove some of the effects of the faster
de-correlating purely damped eigenmodes which are coupled to the
damped oscillatory mode.  Note that due to the coupling of the modes,
filtering cannot completely isolate the behaviour of the oscillatory
mode.  Therefore, predictability even of the filtered process will be
affected by the purely damped modes as well as the damped oscillatory
mode.  The same 10 year low pass filtering that has been applied to
the climatologies shown in Figures \ref{fig:climate_400years} and
\ref{fig:bx4_signal} will be applied to the ensemble elements.  For
the ensemble experiments considered here and in the coupled model, a
10 year low pass filter is quite drastic due to the relatively short
length of the experiments (30 years).  Also, as mentioned in Section
\ref{subsection:model-variability}, filtering increases the e-folding
time of the auto-correlation for a damped stochastic process. It
follows that the predictability of the filtered process will be
increased. In the interests of providing some sense of the added
predictability associated with such a filtering, and since this
particular low pass filter was seen in DMS93 and GT95 to provide a
useful means of hightlighting the low frequency THC variability in the
coupled and box models, we present an analysis of both the yearly
averaged and filtered yearly averaged THC ensemble experiments.

Figure \ref{fig:bx4_ens}A shows the yearly averaged circulation for a
nine element ensemble experiment from the four box model.  The results
are compared in Figure \ref{fig:bx4_ens}B to that from
particular Brownian and harmonic Brownian processes, which are
specified from measurements of the initial position and slope of the
ensemble and parameters estimated from the {\em acf} in Figure
\ref{fig:bx4_signal_acf}.  The ensemble mean relaxes to zero in rough
accordance with both the mean of an infinite ensemble of Brownian and
harmonic Brownian processes, i.e., the damped persistence and damped
harmonic persistence forecasts.  Note, however, that the small
oscillation expected from the damped harmonic persistence forecast is
not manifested in the box model ensemble. Apparently, the damping of
the damped oscillatory eigenmode, and its coupling to the purely
damped processes (recall the system is not self-adjoint), act to
suppress the oscillation expected if the damped oscillatory mode acted
in isolation.

The initial rapid growth in the ensemble variance shown in Figure
\ref{fig:bx4_ens}C agrees with the growth expected from the ensemble
variance in a Brownian process.  This result is interpreted as due to
the presence of the rapidly de-correlating non-oscillatory modes. Also
shown is the ensemble after artificially extending the initial point
10 years to its past and then applying a 10 year low pass filter.  The
closer agreement of the finite sized filtered ensemble statistics with
the damped harmonic persistence solution, especially the initial
growth in ensemble variance, is apparent and can be expected based on
our characterization of the growth in ensemble variance for the
harmonic Brownian process in Section \ref{subsubsection:dho_optimal}
[in particular equation (\ref{eq:optimal_slope})].

\subsection{Four-box summary}
\label{subsection:fourbox_summary}

Although the four-box model contains advective nonlinearities, its
variability as driven by a moderate amplitude white noise forcing can
be understood effectively as that of a system of Brownian processes
coupled to a single harmonic Brownian process.  The coupling arises
from the non-self-adjointness of the linearized system.  The damped
oscillatory eigenmode is responsible for the negative trough in the
{\em acf} shown in Figure \ref{fig:bx4_signal_acf}.  The purely damped
modes and sampling errors in turn act to broaden this trough and the
damping of the oscillation determines the depth of the trough.  The
natural extension of the linear stochastic theory to the present
system of coupled modes also forms a useful theoretical framework for
understanding the predictability of the box model's THC variability.
Of particular note is the rather damped behaviour of the ensemble mean
in which there is no significant oscillatory behaviour.  This result
is consistent with larger ensemble sizes and is the result of the
relatively large damping of the damped oscillatory eigenmode along
with its being coupled to purely damped modes.  Filtering, as
expected, reduces the initial growth of the ensemble variance which
brings it somewhat more in line with the variance from an ensemble of
harmonic Brownian processes.  Box model experiments (not shown) in
which the damped oscillatory eigenmode is less damped, and hence more
dominant in the model's variability, bring out the oscillatory mode in
the ensemble mean and correspondingly increase the predictability of
the circulation.

Some numbers are useful to establish the time scales for
predictability in this model.  The ensemble variance from the yearly
averaged THC, which acts over short time scales much as an ensemble of
red noise processes, reaches a variance $50 \%$ that of the
climatology after roughly 1.5 years.  This time can be used to define
a predictability time scale for the yearly averaged signal from this
model.  The variance of the filtered ensemble reaches $50 \%$ that of
the filtered climatology after roughly 5-7 years.

\section{The coupled model ensemble statistics}
\label{section:coupled-linear}

Following the previous analysis of the four-box model's THC
predictability, the linear stochastic framework is used to help
interpret the statistics from the coupled model ensemble experiments
of Section \ref{section:coupled-model}.  Figure
\ref{fig:ensI_complete} shows the first ensemble experiment from
Figure \ref{fig:ensI} along with the ensemble mean and ensemble
variance.  Both the yearly averaged and 10 year low pass signals are
shown.

The mean of the infinite sized ensemble of Brownian and harmonic
Brownian processes, which initial conditions estimated from the
coupled model ensembles, are superposed with the ensemble mean.  The
decay times and periods for these forecasts were approximated from the
{\em acf}s of the climatology in Figure
\ref{fig:climate_400years_acf}A,B.  Based on our previous experience
with the Brownian and harmonic Brownian processes and their
applications to the box model, Figure \ref{fig:ensI_complete}
indicates that the behaviour of the THC simulated in the coupled model
is consistent with the linear theory much in the same fashion as seen
in the box model results.  The second ensemble, shown in Figure
\ref{fig:ensII_complete}, indicates the same consistency.

The ensemble variance of the yearly averaged signal, which increases
in time similar to the variance from a red noise ensemble, reaches $50
\%$ of the climatological variance after approximately 1.5 years.
This time defines a time scale for the predictability of the yearly
averaged THC index from the coupled model.  The ensemble variance of
the low pass signals reaches $50 \%$ of the climatological variance
after approximately 5-7 years.  This time defines a time scale for the
predictability of the interdecadal THC variability from the coupled
model.

\section{Discussion and conclusions}
\label{section:discussion-conclusions}

A system for monitoring the Atlantic Ocean does not yet exist, but is
under active study by a number of large-scale research programs. This
coupled model study represents a preliminary attempt to assess the
potential of using such an observational system to predict North
Atlantic climate variability on decadal to multi-decadal time scales.
In particular, this investigation reported results of two sets of 30
year ensemble predictability experiments carried out with the GFDL
coupled ocean-atmosphere GCM.  A multi-century integration of the
model was used as a climatology to which to compare the ensemble
statistics.  The ensemble members possess identical oceanic initial
conditions with randomly chosen atmospheric initial conditions taken
from the model climatology.  The motivation for choosing this ensemble
design stems from the hypothesis that it will provide an upper limit
to the model's predictability and hence is useful as a benchmark for
comparing more realistic predictability experiments.  Due to the
connection between the model's North Atlantic climate variability and
its THC, the predictability of the model's THC index variability was
characterized.  This index represents the magnitude of the THC in
the model's North Atlantic.

A case study using a mixed boundary condition four-box model of the
meridional THC motivated a particular linear stochastic intepretation
of the coupled model's THC variability.  This box model, unlike the
prototype model of Stommel (1961), contains a linear damped
oscillatory mode for a stable state corresponding to the present North
Atlantic in which temperature dominates the north-south density
gradient (GT95).  In addition, it contains purely damped modes.  The
damping of the oscillatory mode is related to the strength of the
steady state salinity forcing with a larger forcing bringing the model
into a more oscillatory regime.  The regime corresponding to the
variability of the coupled model is relatively damped.  The box
model's variability can be interpreted with the help of two single
variable linear stochastic processes; namely, the Brownian process
(red noise) and the harmonic Brownian process (noise driven damped
harmonic oscillator).  The model's red noise response corresponds to
the behaviour of the rapidly de-correlating portion of the THC
climatology as indicated its {\em acf} shown in Figure
\ref{fig:bx4_signal_acf}.  The damped harmonic oscillator response
corresponds to the longer time or low frequency behaviour of the THC
indicated by the {\em acf}'s tendency to exhibit a damped oscillatory
response as seen by the presence of a negative trough.  The damping of
the oscillatory behaviour is quite large, however, and the statistical
significance is therefore marginal and must be supported by direct
analysis of the model's linearized dynamics.  Longer time series
support this interpretation.  Since the linearized box model is not
self-adjoint, the model can be thought of as a system of Brownian
processes coupled to a single harmonic Brownian process.

The box model provides a useful case study for applying the basic
ideas of linear stochastic variability and predictability discussed in
Section \ref{section:linear-systems}.  As the statistics from both the
box model and coupled model are quite similar, and based on our belief
that the physical mechanisms responsible for the box model's
variability are related to those seen in a zonally averaged sense from
the coupled model (as argued in GT95), the interpretation of the
coupled model's THC variability and its statistics follows directly
from that of the box model.  This interpretation leads us to believe
that longer ensemble experiments with this model will not alter the
predictability time scales determined in Section
\ref{section:coupled-linear} for the coupled model's THC variability.

The linear stochastic perspective as applied the coupled model's THC
variability suggests that the oceanic variability in the coupled model
is largely driven by atmospheric variability.  Therefore, the model's
THC predictability relies on the ocean's ability to maintain memory
under the effects of this nearly random forcing.  The stochastic
interpretation of this variability is broadly consistent with that
embodied in the stochastic models of Hasselmann (1976,1982),
Mikolajewicz and Meier-Reimer(1990), Weisse et al. (1994), Bryan and
Hansen (1994) and GT95.  The work of Penland and Sardeshmukh (1994)
points to the relevance of such a perspective for ENSO forecasting.

Even with the very long time scales of the coupled model's North
Atlantic climate variability (40--60 years), the rather small
predictability time scale (1.5--7 years) is not unexpected given the
noise driven stochastic interpretation and the rather damped behaviour
of the model's variability.  This situation can be compared to the
roughly one year predictability limit for the 4--7 year time scale
ENSO variability (Neelin et al. 1994).  The large differences in
relative time scales arise from the very different physical mechanisms
acting in the two forms of ocean-atmosphere variability.  However, in
absolute terms the model's THC predictability time scales are somewhat
encouraging from a forecasting perspective, assuming they are not
drastically reduced with more realistic ensemble experiments in which
the oceanic state is perturbed.  Furthermore, although a surface to
deep water monitoring system is probably justified as a means to
monitor global climate change, the present results might provide an
additional justification for such a mid-latitude observing system,
particularly in the North Atlantic.

Although the low order stochastic models discussed in this work are
concise and therefore of value from a conceptual standpoint,
extensions to this work are necessary before it is of practical use
for a forecasting system using a coupled GCM.  Namely, the THC index
is not a readily measurable quantity in the real ocean and as such
should be correlated with an observable field.  As shown by DMS93,
such an observable in the model is the dynamic topography since it is
related through geostrophy to the meridional overturning.  An analysis
similar to that presented here, replacing the THC index time series
with that of the first few dynamic topography EOFs, can be carried
out.  It is expected that the conclusions reached here concerning the
linear noise driven nature of the system would remain unchanged.
However, the utility of understanding the predictability of EOF
patterns would aid in addressing practical issues such as designing an
observational network for the initialization and monitoring necessary
for a forecasting system.  Such a study would be a natural extension
of the current research.  Furthermore, it should be noted that the
model's THC variability exhibits only a moderate amount of variability
($\approx 5\%-10\% $ of the roughly $20$ Sv mean meridional
circulation) as it appears to be quite damped.  Real world
measurements would be hard pressed to capture variability of this
scale.  These limitations hence motivate studies of the variability in
other realistic coupled models in order to understand if this is a
limit to the real coupled ocean-atmosphere system or is somewhat model
dependent.  It should be noted that a severe limitation of the current
coupled GCMs is their course resolution which is far from resolving
oceanic mesoscale eddies.  It remains a topic of future research to
assess the effects these eddies have on long term oceanic
predictability.

{}From an observational perspective, the need to estimate such
climatological statistics as the auto-correlation function of the long
time scale processes simulated here places a heavy burden on the
limited climate data.  The paucity of available data should further
motivate the continuing accumulation of paleoclimate proxy data for
the North Atlantic as well as to motivate the establishment of long
term climate monitoring networks.  With the accumulation of long time
records, validation of models is more available which improves our
confidence in their climatology and ultimately their forecasts.

\appendix

\section{Appendix}
\label{appendix:stationary_stats}

In this appendix, we present the stationary statistics of the Brownian
and harmonic Brownian processes.  More elaboration of these processes
useful for the derivations in this appendix can be found in the books
by Gardiner (1985) and Kubo et al., (1985).

\subsection{The Brownian process}
\label{subsection:app_red_statistics}

The Brownian process is defined by equation (\ref{eq:red}), which is
reproduced here for completeness:
 \begin{equation} \dot{v}(t) = - \alpha v(t) + \xi(t).
 \label{eq:app_red}
 \end{equation}
 A particular solution can be formally written
 \begin{equation}
  v_{p}(t) = \int^{t}_{-\infty} e^{-\alpha(t-u)}\xi(u)du.
 \label{eq:red_solution}
 \end{equation}
 For the subtleties of interpreting the integral of white noise, refer
to Gardiner (1985).  Choosing the lower limit at $-\infty$ for the
particular solution allows the stationary statistics of the process to
be found straightforwardly from this expression.  The stationary mean
$\langle v \rangle$ of the process vanishes.  The stationary
auto-covariance function ({\em acvf}) can be found by multiplying
$v_{p}(t)$ by $v_{p}(s)$ and taking an ensemble mean.  The result for
$s<t$ is
 \begin{equation}
 \langle v(t)v(s)\rangle =
(\sigma_{\xi}^{2}/2\alpha)\exp[-\alpha(t-s)].
 \label{eq:red-acvf}
 \end{equation}
 The covariance between adjacent points in time falls off
exponentially with e-folding time $1/\alpha$.  Note the dependence on
the time difference $(t-s)$; a property characteristic of
statistically stationary processes.  Setting $t=s$ gives the time
independent zero lag variance
 \begin{equation}
 \langle v^{2} \rangle = (\sigma_{\xi}^{2}/2\alpha).
 \label{eq:red-variance}
 \end{equation}
 As expected, the larger the damping coefficient $\alpha$, the smaller
the variance; conversely, the larger $\sigma_{\xi}^{2}$, representing
the power of the noise forcing, the larger the variance.  The
normalized auto-covariance
 \begin{equation}
  \langle v(t) v(s) \rangle \langle v^{2} \rangle^{-1}
  = \exp[-\alpha(t-s)]
 \label{eq:red_correlation}
 \end{equation}
 is called the auto-correlation function({\em acf}).  This damped
exponential has been fit to the {\em acf} for the processes throughout
this paper.

The Fourier transform of (\ref{eq:red}) yields the frequency space
solution $v(\omega) = \xi(\omega)/(\alpha + i \omega)$.  Since the
noise is white, $\xi(\omega) = \sigma_{\xi}$.  The absolute square
$|v(\omega)|^{2}$ gives the spectrum $S(\omega) = 2 \alpha \langle
v^{2} \rangle / (\alpha^{2} + \omega^{2})$.  The form of the spectrum
motivates the name {\em red noise} since the power is concentrated in
the low or red frequency end of the spectrum.  Note that the e-folding
time $\alpha^{-1}$ of the {\em acf}, above which there is basically
no correlation or memory remaining in the process, corresponds to the
angular frequency $\omega = \alpha$, below which the spectrum flattens
out to approximate that of a white noise process with power $2 \langle
v^{2} \rangle \alpha^{-1} = \sigma_{\xi}^{2}/\alpha^{2}$.

\subsection{The harmonic Brownian process}
\label{subsection:app_harmonic_brownian}

The harmonic Brownian process is described by
 \begin{equation}
 \ddot{x}(t) + \omega_{0}^{2} x(t) = -2\beta \dot{x}(t) + \xi(t).
 \label{eq:app_oscillator}
 \end{equation}
 Setting $\beta^{2} < \omega_{0}^{2}$ allows the noise free system to
exhibit a damped harmonic response.  For the opposite inequality, the
motion is overdamped.  Solutions for the overdamped case can be found
from the following underdamped solutions through analytic
continuation.  The noise free solution in equation
(\ref{eq:ho_forecast2}).  This solution is also the ensemble mean of
the process and it vanishes in the stationary limit.  A particular
solution to (\ref{eq:app_oscillator}) is
 \begin{equation}
 x_{p}(t) = \int^{t}_{-\infty} { e^{-\beta(t-u)} \over
        \Omega} \sin[(\Omega(t-u)] \xi(u) du.
 \label{eq:ho_solution}
 \end{equation}
 The {\em acvf} is given by
 \begin{equation}
 \langle x(t)x(s)\rangle  =
  \langle x^{2} \rangle e^{-\beta(t-s)} \{ \cos[\Omega (t-s)]
  + {\beta \over \Omega} \sin[\Omega(t-s)] \},
\label{eq:oscillator_covariance}
\end{equation}
 where the lag $(t-s)>0$.  For $(t-s)<0$, the solution takes the same
form with an absolute value on the lag.  Setting $t=s$ yields the zero
lag variance of the process $\langle x^{2}\rangle = \sigma_{\xi}^{2}
/(4\beta \omega_{0}^{2})$. The {\em acf} $\langle x(t)x(s)\rangle
\langle x^{2} \rangle^{-1}$ is that function which is fit to the
sample {\em acf}s throughout this paper.

 The Fourier transform of the {\em acvf} gives the spectrum
 \begin{equation}
 S_{ho}(\omega) = \langle x^{2} \rangle
        { 4 \beta \omega_{0}^{2}  \over 4\beta^{2} \omega^{2}
         + (\omega_{0}^{2} - \omega^{2})^{2} }.
 \label{eq:ho_spectrum}
 \end{equation}
 The spectrum has a peak value of $\langle x^{2} \rangle \beta^{-1}$
at the angular frequency $\omega = \omega_{0}$, becomes the constant
$4 \beta \langle x^{2} \rangle \omega_{0}^{-2}$ at zero frequency, and
falls off as $\omega^{-4}$ for frequencies $\omega \gg \beta$.  Note
that the spectrum peaks at the oscillator's natural angular frequency
$\omega_{0}$ whereas the {\em acvf} (\ref{eq:oscillator_covariance})
oscillates with the reduced frequency $\Omega =
(\omega_{0}^{2}-\beta^{2})^{1/2}$.

\subsection{Comparison of the stationary statistics}
\label{subsection:comparison}

For the harmonic Brownian process, the high frequency behaviour of the
spectrum and the small time lag behaviour of the {\em acf} are
distinct from the corresponding behaviour of the Brownian process.
Namely, the Brownian process correlation falls off immediately at
non-zero lag time whereas the harmonic Brownian process decays less
rapidly at such lags.  This result has its complementary behaviour in
frequency space with a faster decay at high frequency for the harmonic
process.  The low frequency and long lag time behaviour of the
processes are equivalent as they both approximate the behaviour of
white noise processes in this limit.

The above behaviours manifest in the form of the growth in error
curves for the optimal forecasts of the two processes. Namely, the
error in the damped persistence forecast of the Brownian process has a
nonzero initial slope (equation (\ref{eq:red_optimal_slope}))
corresponding to the sharper drop in small lagged correlation for the
process.  The damped harmonic persistence forecast of the harmonic
Brownian process has a zero initial error slope (equation
(\ref{eq:optimal_slope})) corresponding to the slower initial
de-correlation of the harmonic process.

\vskip 1.5truecm

{\bf Acknowledgments.}

This work would not have been possible without the generous support
and guidance of Tom Delworth, Ron Stouffer, and Eli Tziperman.  We
thank them wholeheartedly.  Further beneficial discussions and advice
were provided by Jeff Anderson, Bhupendra Goswami, Alex Hall, Isaac
Held, Syukuro Manabe, and Cecile Penland.  Thanks also go to GFDL, and
especially its Ocean Group, who kindly provided us with the Cray-YMP
time necessary to conduct this research.  Funding for SMG is provided
by a fellowship from the NOAA Postdoctoral Program in Climate and
Global Change and NOAA's Geophysical Fluid Dynamics Laboratory.
Support was also provided by Atlantic Climate Change Program funding
from NOAA's Office of Global Change.

\newpage

\section{References}

\begin{description}

\item J.L.\ Anderson and W.F.\ Stern, 1995: Evaluating the predictive
utility of ensemble forecasts in a perfect model setting. Preprint.

\item Bjerknes, J.\ 1964: Atlantic air-sea interaction. {\em Advances
in Geophysics}, {\bf 10}, 1-82.

\item Broecker, W.S., D.M.\ Peteet and D.\ Rind, 1985: Does the
ocean-atmosphere system have more than one stable mode of operation?
{\em Nature}, {\bf 315}, 21--26.

\item  Bryan, F., 1986:
High-latitude salinity effects and interhemispheric thermohaline
circulations.  {\em Nature}, {\bf 323}, 301--304.

\item Bryan, K. and F.C. Hansen, 1994: A stochastic model of North
Atlantic climate variability on a decade to century time-scale.  {\em
Proceedings of the Workshop on Decade-to-Century Time Scales of
Climate Variability}, National Research Council, Board on Atmospheric
Sciences and Climate, National Academy of Sciences, Irvine,CA.

\item  Bryan, K. and R. Stouffer, 1991: A note on Bjerknes'
hypothesis for North Atlantic variability.  {\em Journal of Marine
Systems}, {\bf 1}, 229-241.

\item Deser, C. and M.L.\ Blackmon, 1993: Sruface climate variations
over the North Atlantic Ocean during Winter: 1900--1989.
{\em Journal of Climate}, {\bf 6}, 1743--1753.

\item Gardiner, C.W., 1985: {\em Handbook of Stochastic Methods for
Physics, Chemistry, and the Natural Sciences}. Springer-Verlag.

\item Goswami, B. N. and J. Shukla, 1991: Predictability of a coupled
ocean-atmosphere model, {\em Journal of Climate}, {\bf 4}, 3--22.

\item Griffies, S.M. and E. Tziperman, 1995:
A linear thermohaline oscillator driven by stochastic atmospheric
forcing. {\em Journal of Climate} submitted.

\item Hasselmann, K., 1976: Stochastic climate models, part 1--theory.
{\em Tellus}, {\bf 18}, 473-484.

\item Hasselmann, K., 1982: An ocean model for climate variability
studies.  {\em Progress in Oceanography}, {\bf 11}, 69--92.

\item Huang, R.X., J.R. Luyten, and H.M. Stommel, 1992: Multiple
equilibria states in combined thermal and saline circulation.  {\em
J. Phys. Oceanogr.}, {\bf 22}, 231--246.

\item Kleeman, R.\ and S.B.\ Power, 1994: Limits to predictability in
a coupled ocean-atmosphere model due to atmospheric noise. {\em
Tellus}, {\bf 46A}, 529--540.

\item Kloeden, and Platen, 1992: {\em Numerical Solution of Stochastic
Differential Equations}, Springer-Verlag.

\item Kubo, R., M. Toda, and N. Hashitsume, 1985:
{\em Statistical Physics II}, Springer-Verlag.

\item Kushnir, Y., 1994:  Interdecadal variations in North Atlantic
sea surface temperature and associated atmospheric conditions.
{\em Journal of Climate}, {\bf 7}, 141--157.

\item Levitus, S., 1989a: Interpentadal variability of temperature and
salinity at intermediate depths of the North Atlantic Ocean, 1970-1974
versus 1955--1959. {\em Journal of Geophysical Research}, {\bf 94},
6091--6131.

\item Levitus, S., 1989b: Interpentadal variability of salinity in the
upper 150m of the North Atlantic Ocean, 1970-1974
versus 1955--1959. {\em Journal of Geophysical Research}, {\bf 94},
9679--9685.

\item Levitus, S., 1990: Interpentadal variability of steric sea level
and geopotential thickness of the North Atlantic Ocean, 1970-1974
versus 1955--1959. {\em Journal of Geophysical Research}, {\bf 95},
5233--5238.

\item Lorenz, E., 1969: Three approaches to atmospheric predictability.
{\em Bulletin of the American Meteorological Society}, {\bf 50},
345-349.

\item Lorenz, E., 1973: On the existence of extended range
predictability. {\em Journal of Applied Meteorology}, {\bf 12},
543--546.

\item Manabe, S., and R.J. Stouffer, 1988:
Two stable equilibria of a coupled ocean-atmosphere model.
{\em Journal of Climate}, {\bf 1}, 841--866.

\item Manabe, S., and R.J. Stouffer, 1995: Low frequency variation of
surface air temperature in a 1000 year integration of a coupled
ocean-atmosphere model.  Submitted to {\em Journal of Climate}.

\item Manabe, S., R.J. Stouffer, M.J. Spelman, and K. Bryan, 1991:
Transient response of a coupled ocean-atmosphere model to gradual
changes of atmospheric $CO_{2}:$ Part I: Annual mean response. {\em
Journal of Climate}, {\bf 4}, 785--818.

\item Manabe, S., M.J. Spelman, and R.J. Stouffer, 1992: Transient
response of a coupled ocean-atmosphere model to gradual changes of
atmospheric $CO_{2}:$ Part II: Seasonal response. {\em Journal of
Climate}, {\bf 5}, 105--126.

\item Mysak, L.A., T.F. Stocker, and F. Huang, 1993:
Century-scale variability in a randomly forced, two-dimensional
thermohaline ocean circulation model.
{\em Climate Dynamics}, {\bf 8}, 103--116.

\item Mikolajewicz, U., and E. Maier-Reimer, 1990: Internal secular
variability in an ocean general circulation model.  {\em Climate
Dynamics}, {\bf 4}, 145--156.

\item Mikolajewicz, U. and E. Maier-Reiner, 1994:
Mixed boundary conditions in ocean general circulation models and
their influence on the stability of the model's conveyor belt.
{\em Journal of Geophysical Research}, {\bf 99}, 22633--22644.

\item W.H. Munk, 1960: Smoothing and persistence. {\em Journal of
Meteorology}, {\bf 17}, 92--93.

\item Neelin, D.J., M.\ Latif, and F-F.\ Jin, 1994: Dynamics of
coupled ocean-atmosphere models: the tropical problem.  {\em Annual
Review of Fluid Mechanics}, {\bf 26}, 617--659.

\item Penland, C., 1989: Random forcing and forecasting using
principal oscillation pattern analysis. {\em Monthly Weather Review},
{\bf 117}, 2165--2185.

\item Penland, C. and P.D. Sardeshmukh, 1994:
The optimal growth of tropical sea surface temperature anomalies.
{\em Journal of Climate} submitted.

\item Power, S., F.\ Tseitkin, M.\ Dix, R.\ Kleeman, R.\ Colman, and
D.\ Holland, 1995: Stochastic variability at the air-sea interface on
decadal time-scales.  Preprint.

\item Rahmstorf, S., 1994:  Rapid climate transitions in a coupled
ocean-atmosphere model.  {\em Nature}, {\bf 372} 82--85.

\item Reif, F., 1965: {\em Fundamentals of statistical and thermal
physics},  McGraw-Hill.

\item Sausen, R., K.\ Barthel, and K.\ Hasselmann, 1988:
Coupled ocean-atmosphere models with flux correction.
{\em Climate Dynamics}, {\bf 2} 144--163.

\item Schlesinger, M.E.\ and N. Ramankutty, 1994: An oscillation in
the global climate system of period 65--70 years.  {\em Nature}, {\bf
367}, 723--726.

\item Stommel, H., 1961: Thermohaline convection with two stable
regimes of flow.  {\em Tellus}, {\bf 13}, 224--230.

\item Stouffer, R.\ J., S.\ Manabe, and K.\ Bryan, 1989: Interhemispheric
asymmetry in climate response to a gradual increase of atmospheric
$CO_{2}$. {\em Nature}, {\bf 342}, 660-662.

\item Stouffer, R.\ J., S.\ Manabe, and K.\ Ya.\ Vinnikov, 1994:
Model assessment of the role of natural variability in recent global
warming. {\em Nature}, {\bf 367}, 634--636.

\item Tziperman, E. and K. Bryan, 1993: Estimating global air-sea
fluxes from surface properties and from climatological flux data using
an oceanic general circulation model. {\em Journal of Geophysical
Research}, {\bf 98}, 22629--22644.

\item Tziperman, E., R. Toggweiler, Y. Feliks, and K. Bryan, 1994:
Instability of the thermohaline circulation with respect to mixed
boundary conditions: Is it really a problem for realistic models?
{\em Journal of Physical Oceanography}, {\bf 24}, 217--232.

\item Weaver, A.J. and T.M.C. Hughes, 1992: Stability and variability
of the thermohaline circulation and its link to climate. {\em Trends
in Physical Oceanography}, {\bf 1}, 15--70.

\item Weisse, R., U. Mikolajewicz, and E. Maier-Reimer, 1994: Decadal
variability of the North Atlantic in an ocean general circulation
model.  {\em Journal of Geophysical Research}, {\bf 89}, 12411--12421.

\item Welander, P., 1986: Thermohaline effects in the ocean
circulation and related simple models.  {\em Large-Scale Transport
Processes in the Oceans and Atmosphere}, D.L.T. Anderson and J.
Willebrand, Eds., NATO ASI series, Reidel.

\item Zhang, S., R.J. Greatbatch, and C.A. Lin, 1993: A re-examination
of the polar halocline catastrophe and implications for coupled
ocean-atmosphere models. {\em Journal of Physical Oceanography}, {\bf
23}, 287--299.

\end{description}

\newpage

\centerline{\psfig{figure=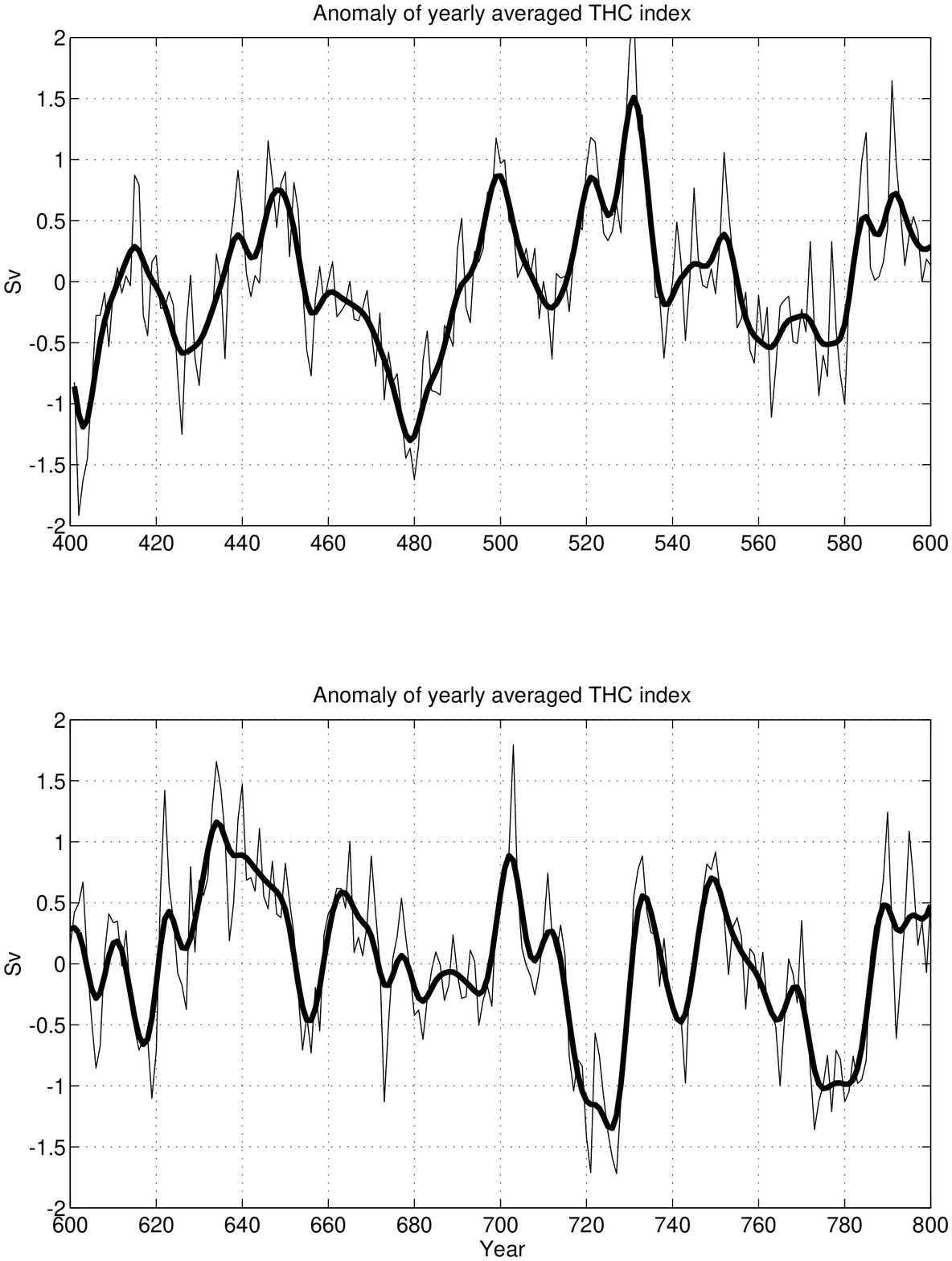,height=5.0in}}
\begin{figure}[htbp] \caption{
  \baselineskip 3ex
  400 years of linearly detrended anomalous yearly averaged
  THC index from the central portion of the coupled model experiment
  of DMS93 (thin solid line).  This time series defines
  the climatology for use in quantifying the predictability
  exhibited by the ensemble experiments.
  The mean (subtracted out) and standard deviation are
  18.1 Sv and .66 Sv (1 Sv = $10^{6} {\mbox m}^{3}/{\mbox sec}$),
  respectively. Also shown is the time series
  after a 10 year low pass filtering has been applied (thick solid line).
  The standard deviation of the filtered time series is .54 Sv.}
\label{fig:climate_400years}
\end{figure}

\centerline{\psfig{figure=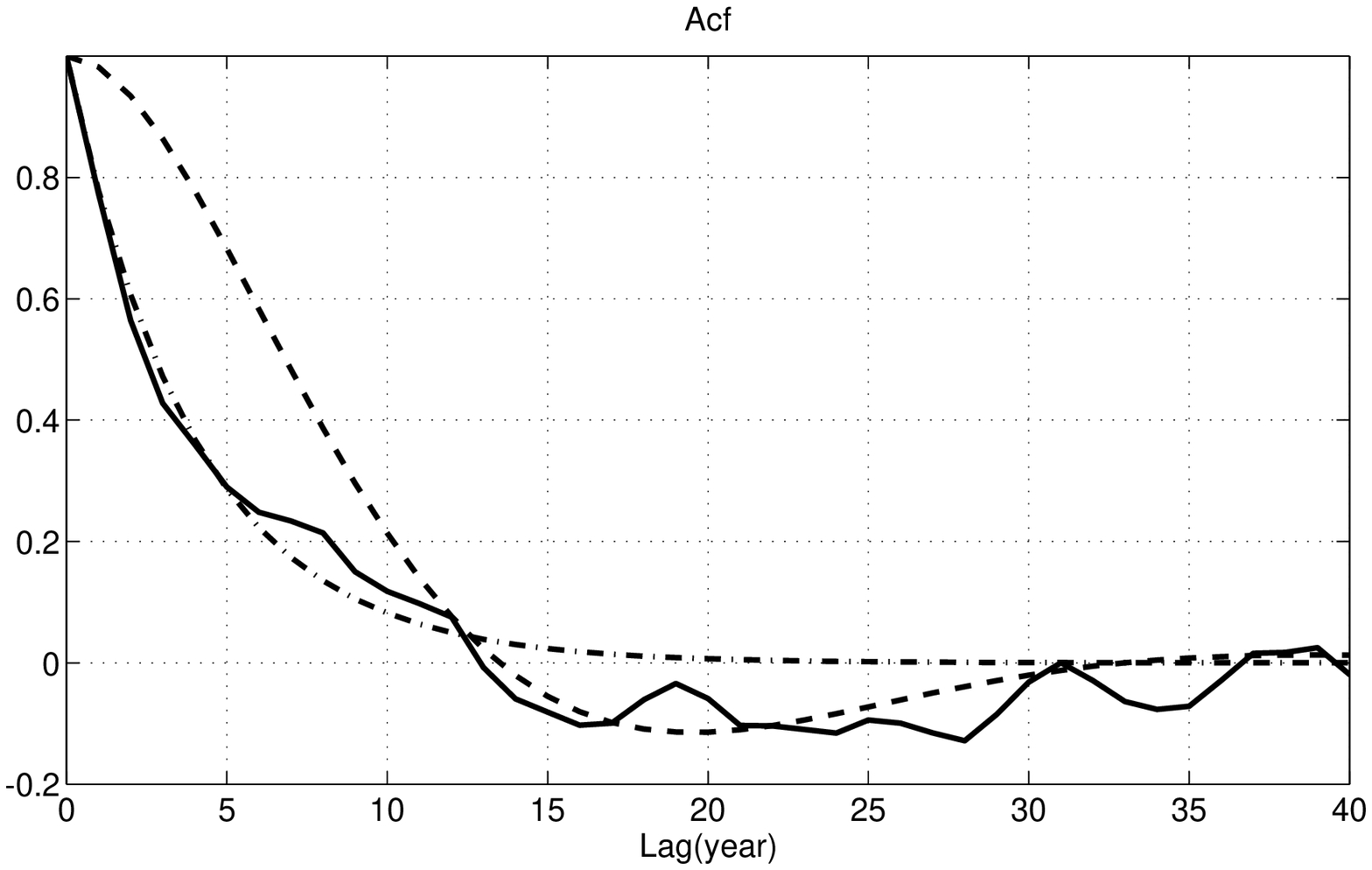,height=2.5in}}
\begin{figure}[htbp] \caption{
  \baselineskip 3ex
  Auto-correlation function ({\em acf}) for the THC index
  shown in Figure \protect\ref{fig:climate_400years}.  The
  smooth curves are fits of the {\em acf}
  from the Brownian or red noise process (dot-dashed line)
  with e-folding time of 4 years to the lags $\le 10$ years, and
  the {\em acf} from the harmonic Brownian process (dashed
  line) with period and e-folding time 31 and 9 years, to lags
  $\ge 10$ years.
  A similar fit (not shown) to the {\em acf} of the low pass THC index
  time series gives a 7 year
  e-folding for the Brownian process and period and e-folding of 35
  and 11 years, respectively, for the harmonic Brownian process. }
\label{fig:climate_400years_acf}
\end{figure}

\newpage

\centerline{\psfig{figure=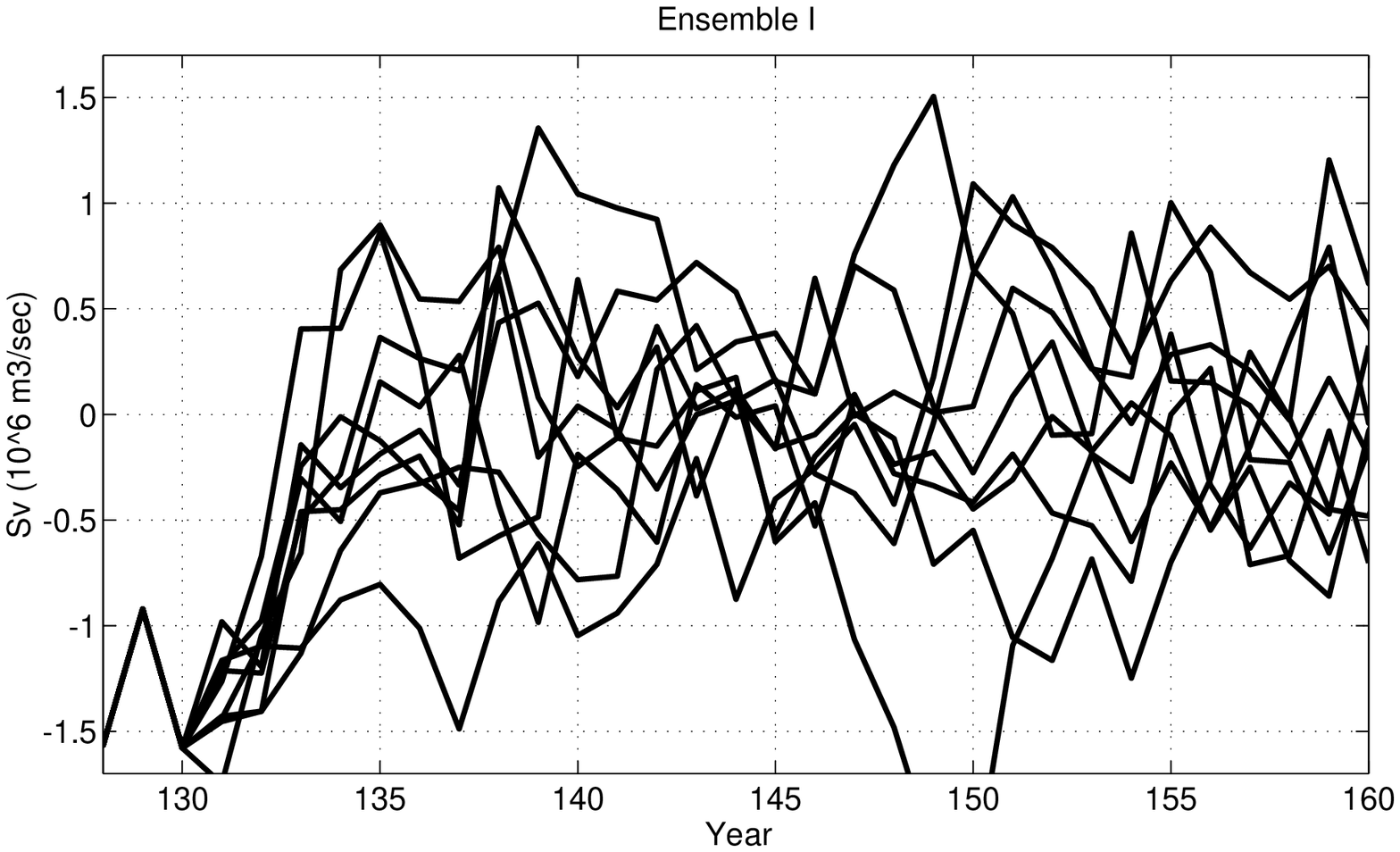,height=3.0in}}
\begin{figure}[htbp]
\caption{
   \baselineskip 3ex
  Nine element ensemble of yearly
  averaged anomalous THC index from the coupled model.
  Anomalies are defined relative to the mean of the linearly
  detrended 400 year climatology of Figure
  \protect\ref{fig:climate_400years}A. The ensemble starts at
  year 130 and extends for 30 years. }
\label{fig:ensI}
\end{figure}

\centerline{\psfig{figure=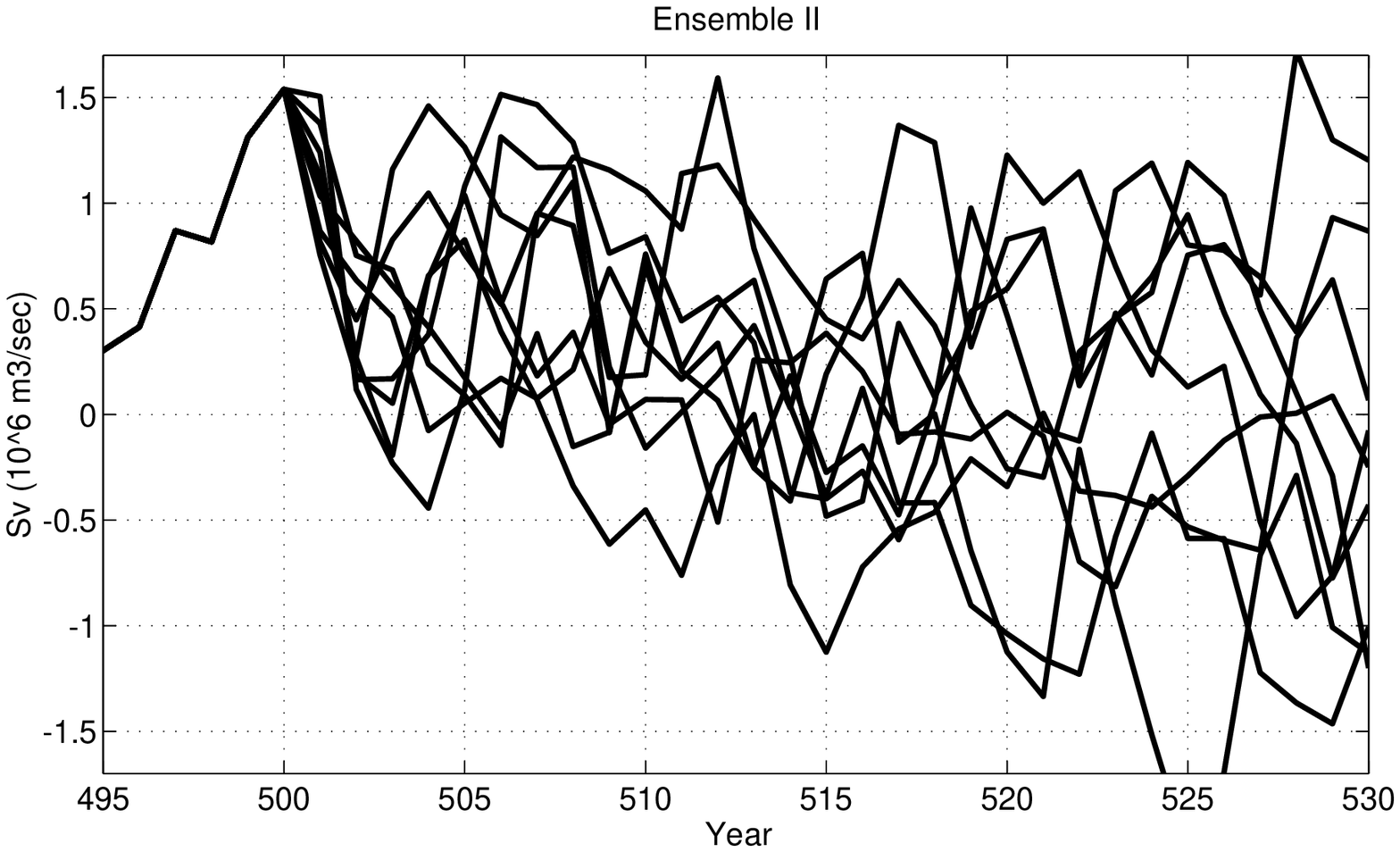,height=3.0in}}
\begin{figure}[htbp]
\caption{
   \baselineskip 3ex
   Same as Figure \protect\ref{fig:ensI} for the ensemble
   starting from year 500 of the climatology.}
\label{fig:ensII}
\end{figure}

\centerline{\psfig{figure=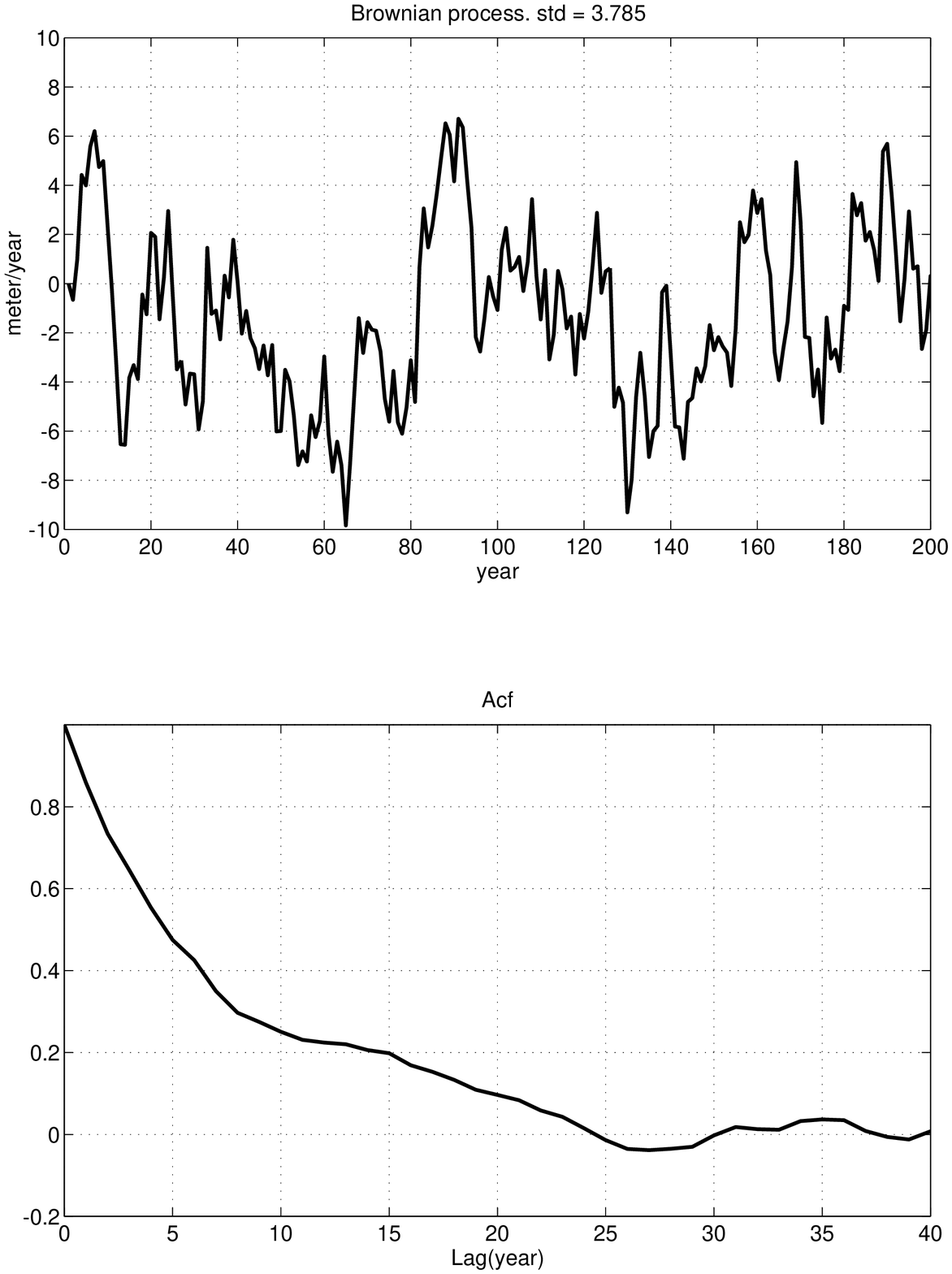,height=5.0in}}
\begin{figure}[htbp]
\caption{
  \baselineskip 3ex
  ({\bf A}) A 200 year realization of the Brownian process
  [the continuous process is described by equation (\protect\ref{eq:red})]
  using an inverse damping time $\alpha^{-1} = 7$ years and noise
  variance $4 \, \mbox{meter}^{2}/\mbox{years}^{3}$.
  The standard deviation of the process in the continuous case
  is $3.7 \, \mbox{meter} / \mbox{year}$, and that realized in the
  numerical case is indicated on the figure title.
  ({\bf B}) The {\em acf} of 400 years of the Brownian process,
  including the 200 years in ({\bf A}).  This function approximates
  that from the infinite sized ensemble, which is a pure exponential
  decay, given by equation (\protect\ref{eq:red-acvf}) in the Appendix.
  An $\approx 7$ year e-folding time is apparent.  Note that for
  illustrative purposes, we chose to set the dimensions of the process
  $v(t)$ to be those of a velocity, as is relevant for Brownian motion.}
\label{fig:Brown_corr}
\end{figure}

\centerline{\psfig{figure=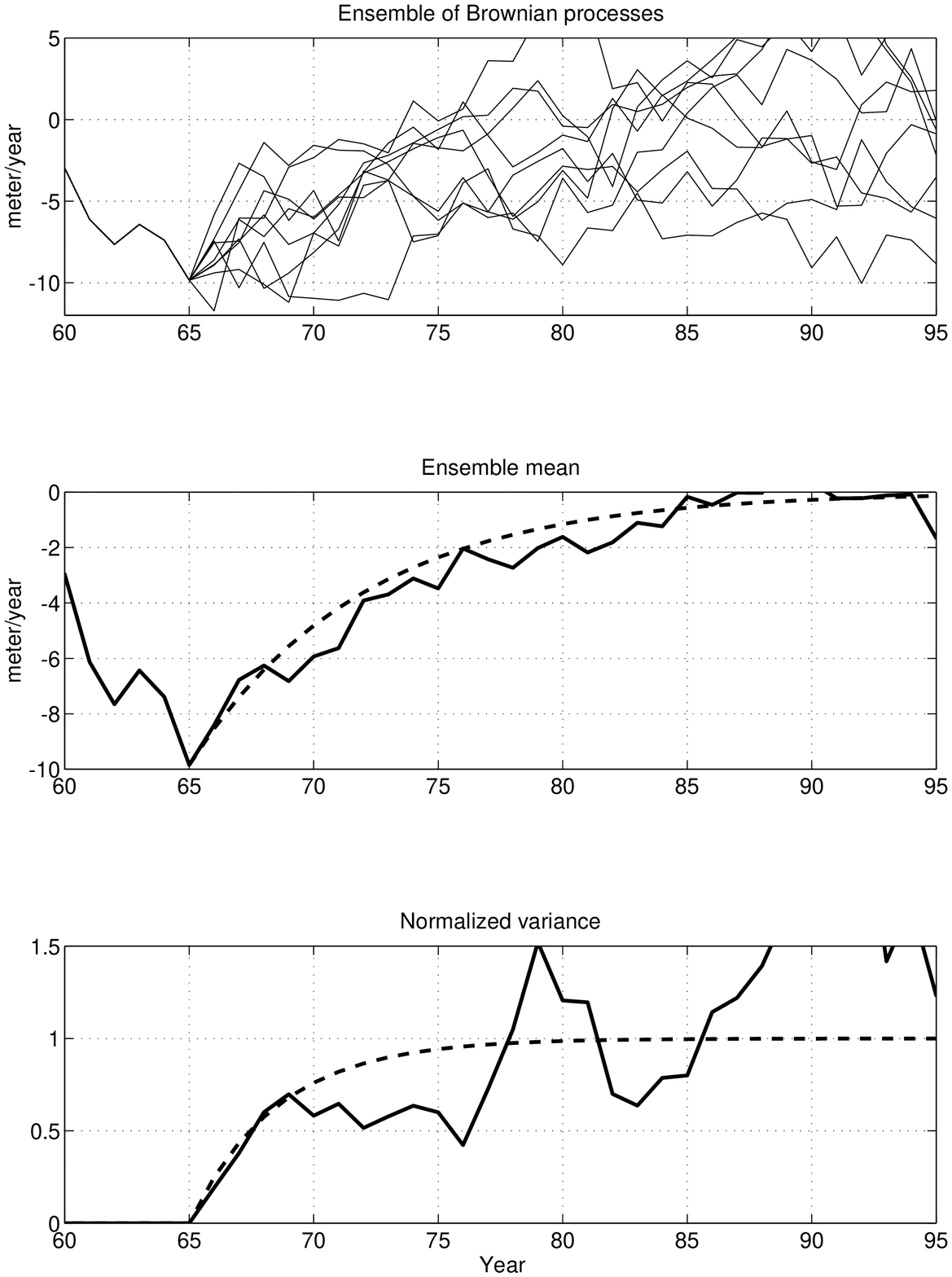,height=6.0in}}
\begin{figure}[htbp]
\caption{
  \baselineskip 3ex
  ({\bf{A}}) A nine element ensemble of the Brownian process,
  each element of which has a different realization of the
  white noise process after starting from the same initial state
  near year 65.
  ({\bf{B}}) The ensemble mean (solid line) and fit of the damped
  persistence [equation (dashed line;
  equation (\protect\ref{eq:red_forecast})].
  Note the different vertical axis relative to ({\bf{A}}).
  ({\bf{C}})  The ensemble variance (solid line) and the error function
  [dashed line; equation (\protect\ref{eq:red_sigma})] of the
  damped persistence forecast.
  Note the non-zero slope of the variance at zero lag consistent with
  equation (\protect\ref{eq:red_optimal_slope}).   }
\label{fig:Brown_ensemble}
\end{figure}

\centerline{\psfig{figure=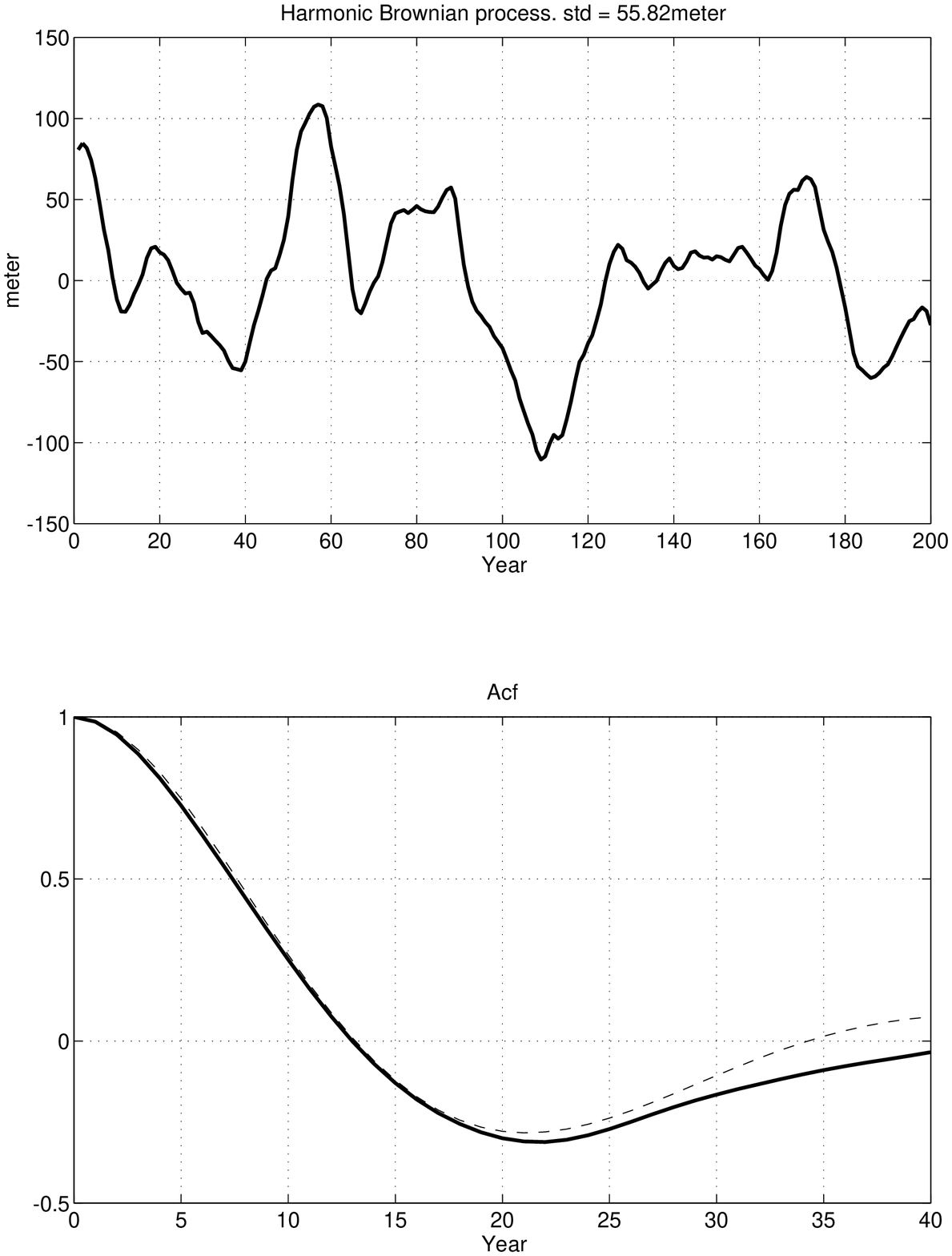,height=4.0in}}
\begin{figure}[htbp]
\caption{
  \baselineskip 3ex
  ({\bf A}) A 200 year realization of the harmonic Brownian
  process [the continuous process is described by equation
  (\protect\ref{eq:oscillator})] using an inverse damping time
  $\beta^{-1} = 10$ years, period $2\pi/\omega_{0} = 33$ years.
  The standard deviation of the harmonic
  Brownian process in the continuous case
  is $52.5$ meter.
  The realized standard deviation is indicated on the figure title.
  ({\bf B}) The {\em acf} for 400 years of the harmonic Brownian
  process (solid line), including the 200 years of ({\bf A}).
  A least squares fit of the theoretical {\em acf}
  (equation (\protect\ref{eq:oscillator_covariance}) discussed
  in the Appendix) to the sample {\em acf} is
  the dashed line.  The period and decay time from this
  fit are 39 and 17 years, respectively.
  Note the broadening of the trough relative to the theoretical {\em acf}
  can be attributed to sampling errors. The depth of the trough is related to
the
  damping of the oscillation (more damping yields a shallower trough).}
\label{fig:harm_Brown_corr}
\end{figure}

\centerline{\psfig{figure=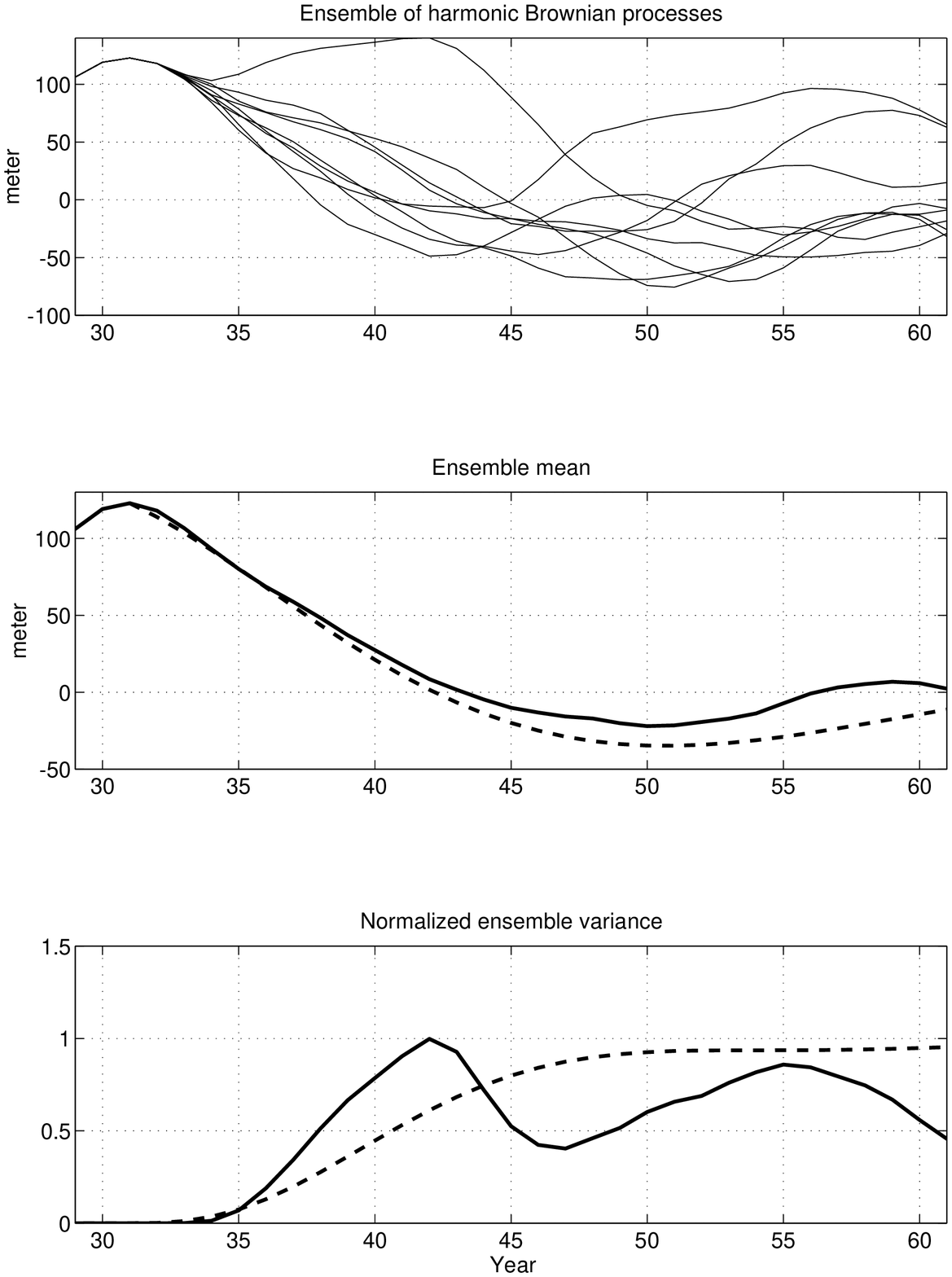,height=5.0in}}
\begin{figure}[htbp]
\caption{
  \baselineskip 3ex
  ({\bf A}) A nine element ensemble of the harmonic Brownian
  process.  Each element was integrated for
  an initial period using the same
  white noise forcing.  Afterwards, the noise is unique to each
  element.
  ({\bf B}) The ensemble mean (solid) and the damped harmonic
  persistence forecast (equation (\protect\ref{eq:ho_forecast2});
  dashed line).
  ({\bf C})  The ensemble variance (solid) and the error function
  of the damped harmonic persistence forecast
  (equation (\protect\ref{eq:ho_forecast_error}); dashed line).
  The parameters for these forecasts were estimated from the
  {\em acf} of Figure \protect\ref{fig:harm_Brown_corr}.
  Note the approximately zero initial slope in the variance
  representing the slower initial de-correlation of the ensemble elements
  relative to the Brownian process ensemble shown in Figure
  \protect\ref{fig:Brown_ensemble}. }
\label{fig:harm_Brown_ensemble}
\end{figure}

\centerline{\psfig{figure=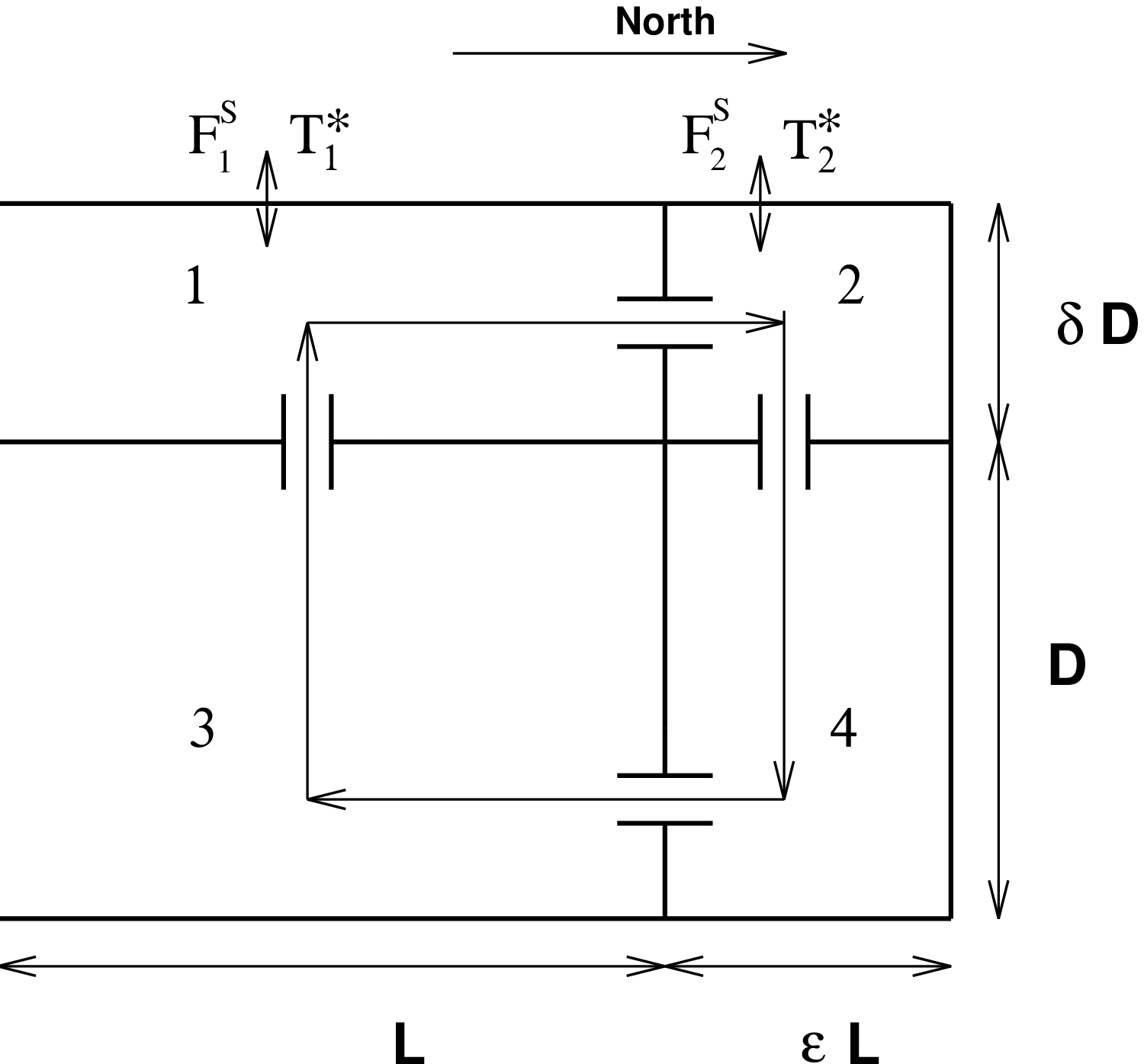,height=5.0in}}
\begin{figure}[htbp]
\caption{
   \baselineskip 3ex
   Configuration of the four box model. The thermal driven
   mean circulation, with sinking in the north and rising in
   the south, is indicated.  The boxes are homogeneous.
   The parameters chosen for
   the numerical experiments are the following:
   $V=8 \times 10^{16}{\mbox m}^{3},
   \epsilon=.10, \delta=.10$, and  $D=3000 {\mbox m}$.
   The surface box temperatures are restored to
   $T_{1}^{*}=25^{\circ}\mbox{C}$ and $T_{2}^{*}=0^{\circ}\mbox{C}$
   with a restoring time $\gamma_{T}^{-1} = 180$ days, which
   corresponds to a restoring time of .6 day for each metre depth.
   The surface salinity forcing
   $F^{S}_{1,2}$ is restoring to the salinities $S_{1}^{*} = 36.5$ psu
   and $S_{2}^{*} = 34.5$ psu using $\gamma_{S} = .2\gamma_{T}$
   ($\gamma_{S}^{-1} = 900$ days) during the integration to steady state.
   Afterwards, a fixed salinity flux
   $F^{S}_{2} = -\epsilon F^{S}_{1} =
   \gamma_{S}(S^{*}_{2}-\overline{S}_{2}) < 0$
   is used for the mixed boundary condition integrations.
   Note the four boxes all have different volumes.}
\label{fig:4box_geometry}
\end{figure}

\centerline{\psfig{figure=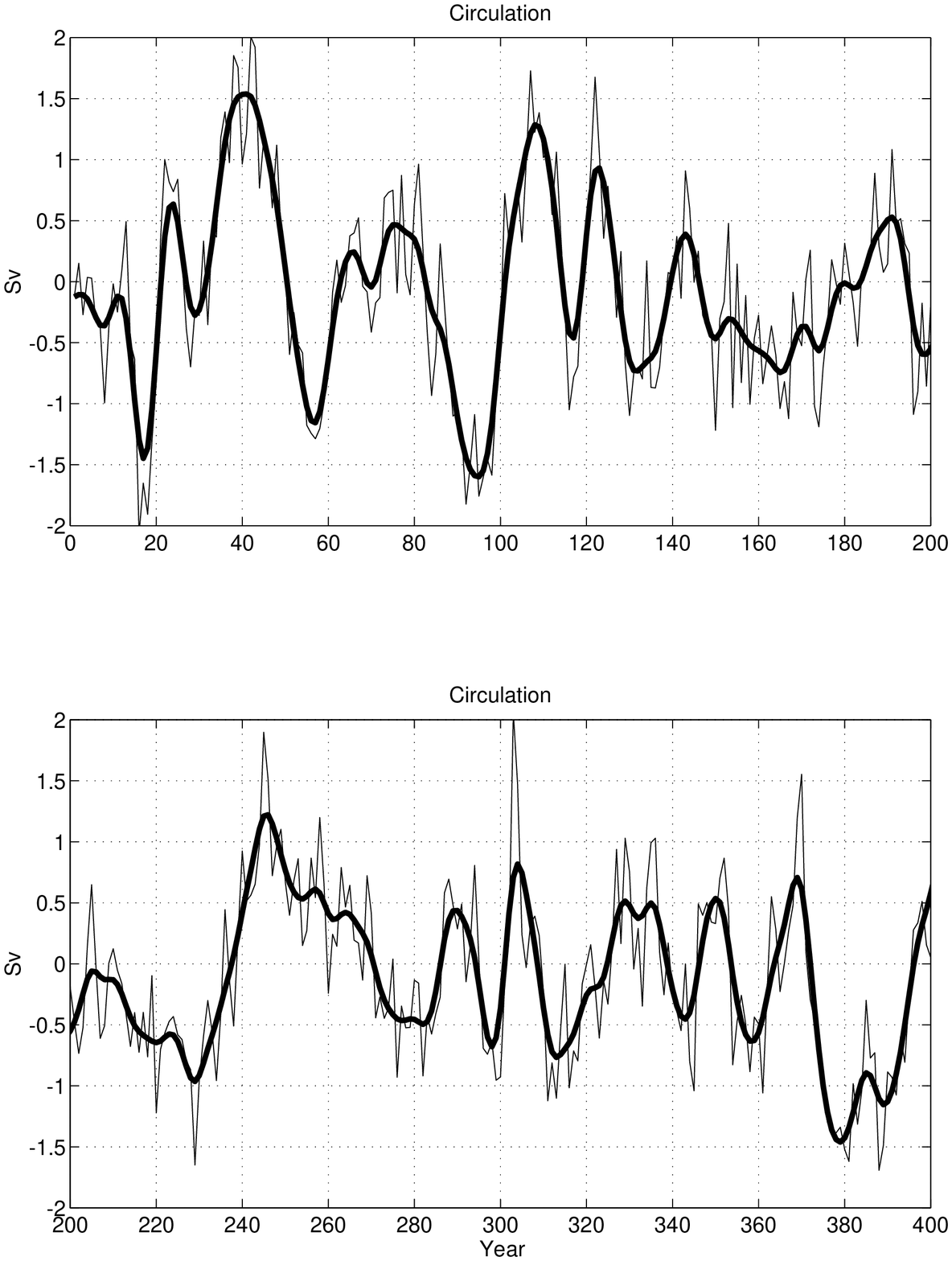,height=5.0in}}
\begin{figure}[htbp]
\caption{
   \baselineskip 3ex
   A 400 year realization of the yearly averaged
   circulation in the four box model (thin solid line).
   The mean (which has been subtracted) is 19.6 Sv and the
   standard deviation is .76 Sv.
   A 10 year low pass filtered (thick solid line) version
   is also shown.  The standard deviation of the filtered
   time series is .65 Sv.  The variability seen in this
   time series is meant to roughly correspond to that of the THC
   index generated by the coupled model seen in
   Figure \protect\ref{fig:climate_400years}. }
\label{fig:bx4_signal}
\end{figure}

\newpage

\centerline{\psfig{figure=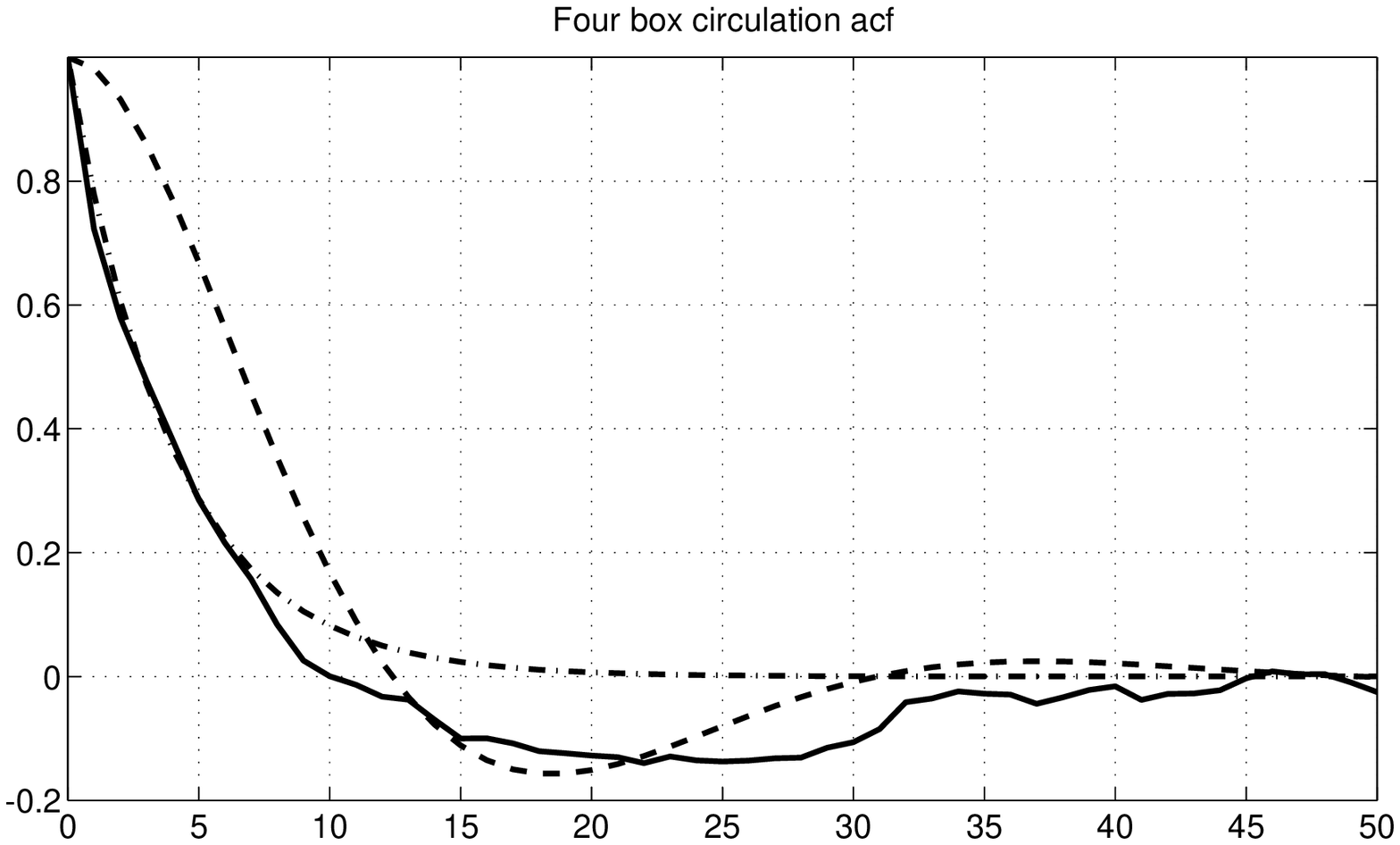,height=2.5in}}
\begin{figure}[htbp]
\caption{
  \baselineskip 3ex
   {\bf (A)} {\em Acf} of the circulation shown in
   Figure \protect\ref{fig:bx4_signal}.  The damped exponential
   (dot-dashed line) is the {\em acf} for a Brownian process
   with e-folding time $4$ years.
   This {\em acf} matches that of the box model at small
   lag times reflecting the small time scale exponentially
   damped modes causing de-correlation at this scale.
   The damped oscillatory function (dashed
   line), which matches the {\em acf}
   at longer lag times where there is a trough due to
   the oscillatory eigenmode, is the {\em acf} for a harmonic
   Brownian process. The period and decay times of this process are
   32  and  10 years, respectively.
   A similar fit to the low pass climatology (not shown)
   gives an e-folding of 6 years for the Brownian process and
   a period and e-folding time of 35 and 12 years, respectively
   for the harmonic Brownian process.  Compare to the {\em acf}
   for the coupled model's THC index shown in Figure
   \protect\ref{fig:climate_400years_acf}.  }
\label{fig:bx4_signal_acf}
\end{figure}

\newpage

\centerline{\psfig{figure=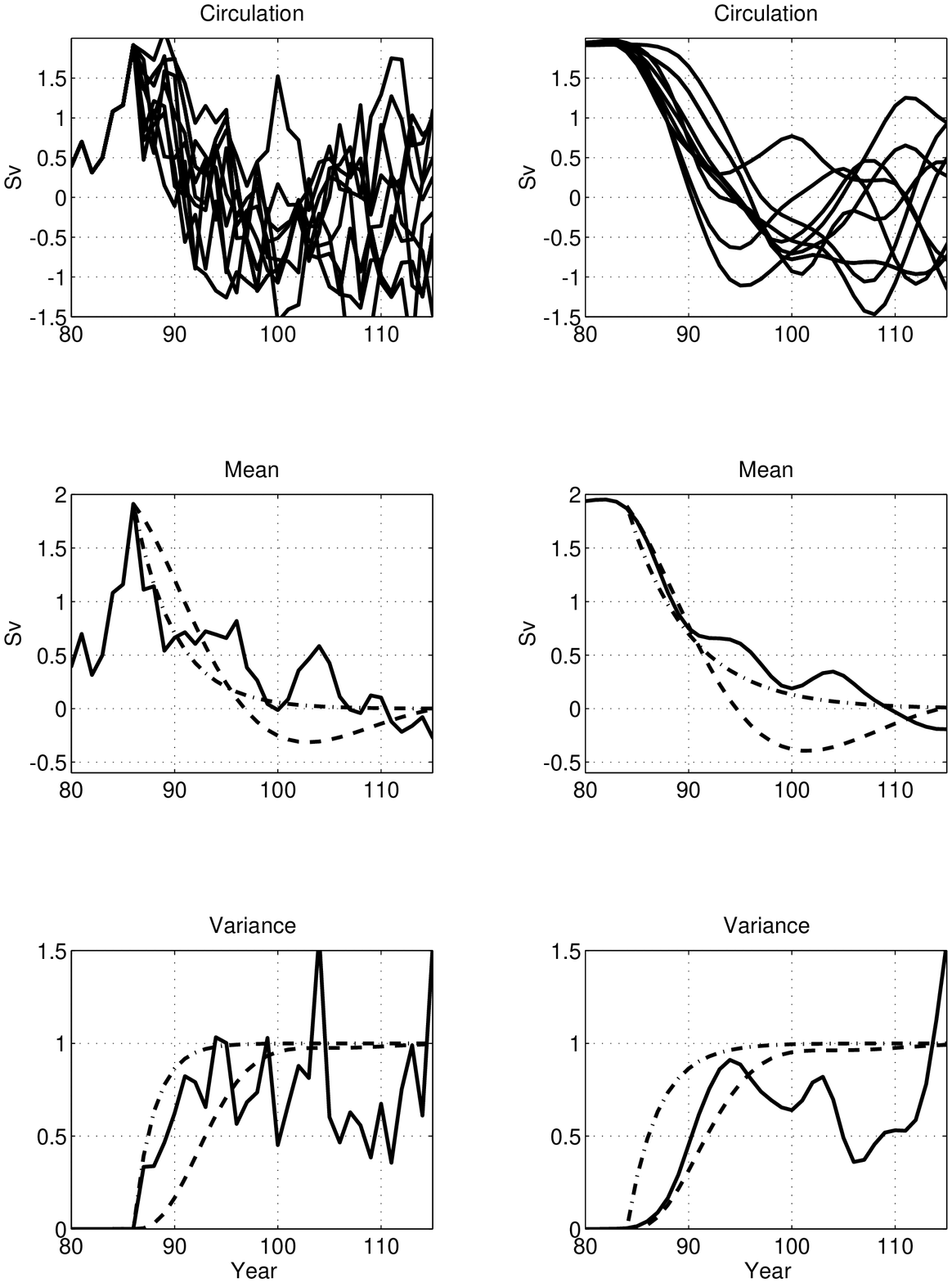,height=5.0in}}
\begin{figure}[htbp]
\caption{
   \baselineskip 3ex
   {\bf (A)} A nine element ensemble of yearly averaged circulation
   from the four box model.
   {\bf (B)} Ensemble mean(solid line) as well as the
   mean from an infinite ensemble of Brownian processes
   (i.e., the damped persistence forecast
   (\protect\ref{eq:red_forecast}); dot-dashed line)
   and the mean from an infinite
   ensemble of harmonic Brownian processes
   (i.e., the damped harmonic persistence forecast
   (\protect\ref{eq:red_forecast}); dashed line)
   using parameters estimated from fits to the
   climatology in Figure \protect\ref{fig:bx4_signal_acf}.
   {\bf (C)} Ensemble variance (solid line).
   Also shown are the variances from the ensemble of Brownian
   processes (i.e., the error functions from the
   damped persistence forecast
   (\protect\ref{eq:red_sigma});  dot-dashed line)
   and that from the ensemble of harmonic Brownian processes
   (i.e, the damped harmonic persistence forecast error
   (\protect\ref{eq:ho_forecast_error}); dashed line).
   Note the rapid initial growth (slope $\ne 0$)
   in variance corresponding to the damped persistence forecast.
   {\bf (D), (E), (F)} are the same as {\bf (A), (B), (C)} after
   a 10 year low pass filtering is applied.  The initial
   point is extended 10 years prior to the ensemble's start
   to facilitate low pass filtering.
   The slower growth in initial ensemble variance resulting
   from the filtering bringing it somewhat
   more in line with the variance
   from the harmonic Brownian process.}
\label{fig:bx4_ens}
\end{figure}

\newpage

\centerline{\psfig{figure=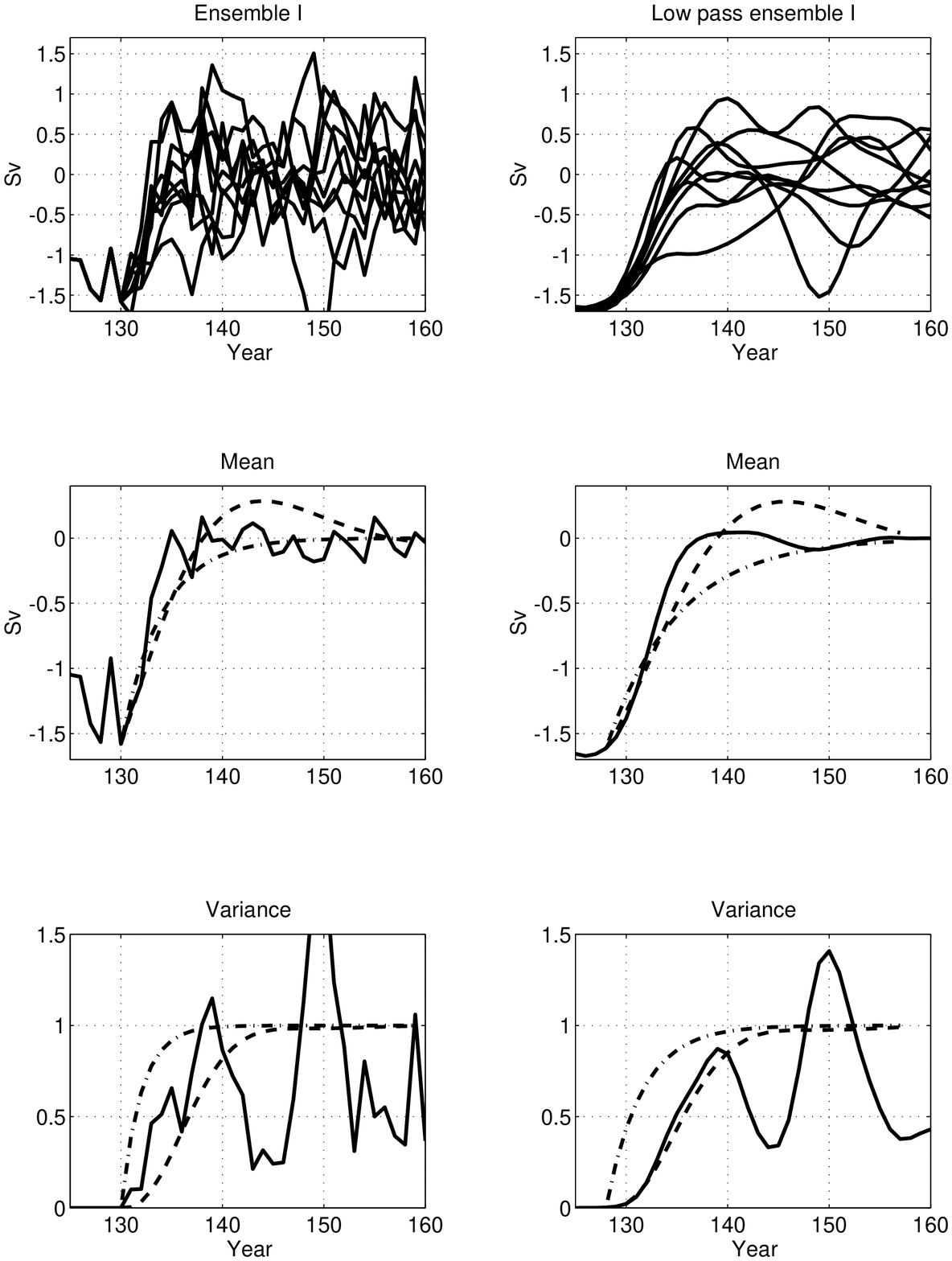,height=5.0in}}
\begin{figure}[htbp]
\caption{
   \baselineskip 3ex
  {\bf (A)} Nine element ensemble of yearly
  averaged anomalous THC index from the coupled model.
  The ensemble starts at year 130 and extends for 30 years.
  {\bf (B)} The ensemble mean.  Also shown are the damped persistence
  forecasts (dot-dashed) and damped harmonic persistence forecasts
  (dashed) derived from the fit of the Brownian and
  harmonic Brownian processes to the 400 year climatology of
  Figure \protect\ref{fig:climate_400years_acf}.
  {\bf (C)} The normalized ensemble variance.
  {\bf (D)} Same as  {\bf (A)} for the 10 year low pass signals
  where the anomalies are defined relative to the climatology
  indicated in the caption of Figure \protect\ref{fig:climate_400years}.
  {\bf (E)} Same as {\bf (B)} for the low pass ensemble.
  {\bf (F)} Same as {\bf (C)} for the low pass ensemble.   }
\label{fig:ensI_complete}
\end{figure}

\newpage

\centerline{\psfig{figure=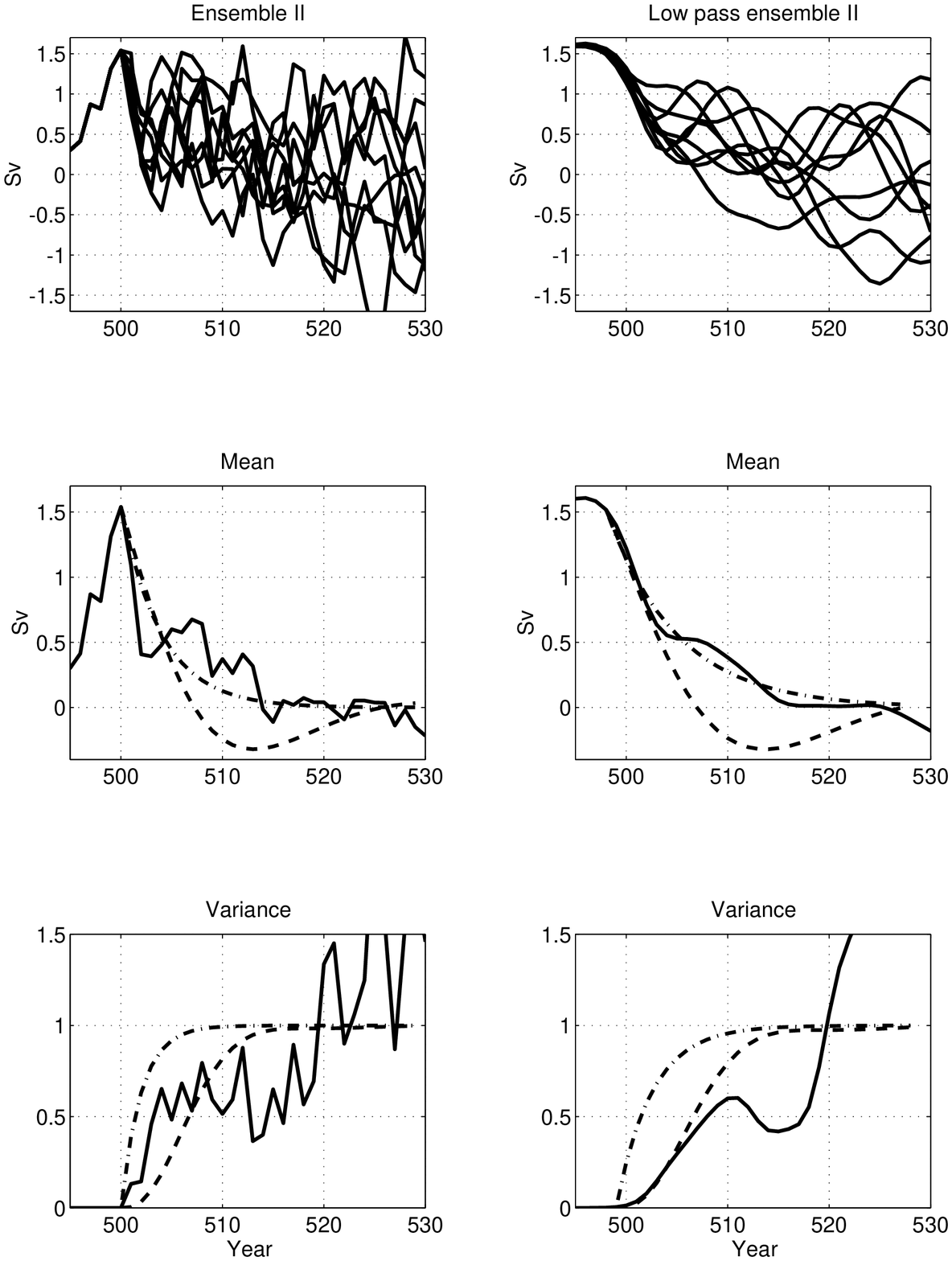,height=5.0in}}
\begin{figure}[htbp]
\caption{
 \baselineskip 3ex
 {\bf (A)--(F)}.  Same as Figure \protect\ref{fig:ensI_complete}
 {\bf (A)--(F)} for the second coupled model ensemble experiment.}
\label{fig:ensII_complete}
\end{figure}

\end{document}